\documentclass[aps,prev,twocolumn,preprintnumbers,floatfix,nofootinbib]{revtex4-1}

\usepackage[T1]{fontenc} 
\usepackage[utf8]{inputenc}
\usepackage{physics}

\usepackage[dvipdfm]{graphicx}
\usepackage{mathrsfs}
\usepackage{amssymb}
\usepackage{verbatim}
\usepackage{color}
\usepackage[dvipsnames]{xcolor}
\usepackage{multirow}
\usepackage{amsmath}
\usepackage{subfig}
\usepackage{slashed} 
\usepackage{float}
\usepackage[normalem]{ulem}
\usepackage{sidecap}
\usepackage{hyperref}
\usepackage{cancel}
\usepackage{enumerate}
\hypersetup{pdfstartview=FitV,colorlinks=true,linkcolor=blue,citecolor=red,filecolor=black,urlcolor=blue}

\newcommand*\diff{\mathop{}\!\mathrm{d}}

\newcommand{\aend}{a_{\rm end}}
\newcommand{\arh}{a_{\rm RH}}
\newcommand{\rhorh}{\rho_{\rm RH}}
\newcommand{\rhoend}{\rho_{\rm end}}
\newcommand{\beq}{\begin{equation}}
\newcommand{\eeq}{\end{equation}}
\newcommand{\bea}{\begin{eqnarray}}
\newcommand{\eea}{\end{eqnarray}}

\def\trh{T_{\rm RH}}
\def\tmax{T_{\rm max}}

\def\del{\partial}
\def\lrd{\overset{\leftrightarrow}{\del}}
\def\ld{\overset{\leftarrow}{\del}}

\newcommand{\lrnabla}{\overset{\leftrightarrow}{\nabla}}
\newcommand{\lnabla}{\overset{\leftarrow}{\nabla}}
\newcommand{\rnabla}{\overset{\rightarrow}{\nabla}}

\usepackage[parfill]{parskip}
\begin{document}\sloppy

 \preprint{UMN--TH--4225/23, FTPI--MINN--23/17, OU--HET--1204}

\vspace*{1mm}

\title{Gravitational Production of Spin-3/2 Particles During Reheating}
\author{Kunio Kaneta$^{a}$}
\email{kaneta@het.phys.sci.osaka-u.ac.jp}
\author{Wenqi Ke$^{b,d}$}
\email{wke@umn.edu}
\author{Yann Mambrini$^{c}$}
\email{yann.mambrini@ijclab.in2p3.fr}
\author{Keith A. Olive$^{d}$}
\email{olive@umn.edu}
\author{Sarunas Verner$^{e}$}
\email{verner.s@ufl.edu}
\vspace{0.5cm}
\affiliation{${}^a $ Department of Physics, Osaka University, Toyonaka, Osaka 560-0043, Japan}
\affiliation{${}^b$ Sorbonne Universit\'e, CNRS, Laboratoire de Physique Th\'eorique et Hautes Energies, LPTHE, F-75005 Paris, France.
}
\affiliation{${}^c $  Universit\'e Paris-Saclay, CNRS/IN2P3, IJCLab, 91405 Orsay, France}
\affiliation{${}^d$William I. Fine Theoretical Physics Institute, School of
 Physics and Astronomy, University of Minnesota, Minneapolis, MN 55455,
 USA}
 \affiliation{${}^e$Institute for Fundamental Theory, Physics Department, University of Florida, Gainesville, FL 32611, USA}

\date{\today}

\begin{abstract} 
We compute the density of a spin-$\frac32$ particle, the raritron, produced at the end of inflation due to gravitational interactions. We consider a background inflaton condensate as the source of this production, mediated by the exchange of a graviton. This production
greatly exceeds the gravitational production from the emergent thermal bath during reheating. The relic abundance limit sets an absolute minimum mass for a stable raritron, though there are also model dependent constraints imposed by unitarity. We also examine the case of gravitational production of a gravitino, taking into account the goldstino evolution during reheating. We compare these results with conventional gravitino production mechanisms. 
\end{abstract}

\maketitle

\setcounter{equation}{0}

\section{Introduction}
Any inflationary theory consists of three key components~\cite{reviews}. First, it must have a prolonged period of exponential expansion to account for the observed flatness of the Universe. Second, the produced density fluctuations should agree with the CMB measurements of the anisotropy spectrum and tensor-to-scalar ratio~\cite{Planck}. Finally, the theory should incorporate a reheating phase, resulting in a hot thermal universe. This universe should have a minimum temperature of a few MeV to allow Big Bang Nucleosynthesis (BBN), and potentially even a higher temperature nearing the TeV scale or higher, which is necessary for baryogenesis. 

Reheating is most efficient when a direct decay channel exists for the inflaton to Standard Model (SM) fields~\cite{dg,nos}. Assuming that the decay products thermalize instantly and with an inflaton potential $V(\phi)$ which is quadratic about its minimum, the reheating temperature is directly related to the 
inflaton decay rate, $\trh \propto (\Gamma_\phi M_P)^\frac12$, where $\Gamma_\phi$ is the decay rate for the inflaton, $\phi$, and $M_P= 1/\sqrt{8 \pi G_N}\simeq 2.4\times 10^{18}$ GeV is the reduced Planck mass. 
The reheating process is not instantaneous; at the end of inflation, the inflaton decays producing a bath of relativistic particles~\cite{Giudice:2000ex,Bernal:2020gzm,GMOP}.
When an inflaton potential is predominantly characterized by a quadratic term near its minimum, the inflaton energy density scales as $\rho_\phi \sim a^{-3}$, where $a$ is the cosmological scale factor. The radiation density rapidly increases, reaching a peak temperature, $\tmax$, 
which then falls until the energy density in radiation becomes equal to that stored in the inflaton 
condensate, thus defining the reheating temperature.
The reheating temperature and the scaling of the radiation density, in principle, depend on the spin of the final state particle and the shape of the potential 
near the minimum that governs inflaton oscillations \cite{GKMO1,GKMO2,Becker:2023tvd}. 

Once produced, the thermal bath can generate very weakly coupled non-SM particles that do not 
achieve thermal equilibrium \cite{fimp,Mambrini:2013iaa}.
Importantly, these might include a dark matter component.
The gravitino is a classic example of such a feebly interacting massive particle, or FIMP~\cite{nos,ehnos,kl,ekn,oss}. For a 
review and related studies, see \cite{Bernal:2017kxu,Bhattacharyya:2018evo,Bernal,Bernal:2019mhf,Barman:2020plp,Chen:2017kvz}. The relic density of a FIMP is determined by its thermally averaged production cross section from the thermal bath. Consequently, the relic abundance is sensitive to $\trh$ or $\tmax$, which depends on the form of its coupling to the SM. 

In addition to the production of 
matter and dark matter from the thermal bath, it 
is also possible to produce matter directly from
inflaton decays or scatterings in which case the relic density depends on 
the coupling of the matter to the inflaton 
\cite{egnop,grav2,GMOP,GKMO1}. 
In the absence of a direct coupling 
between the inflaton and dark matter, 
radiative decays of the inflaton may produce a significant relic density~\cite{KMO,moz}, provided the dark matter has a coupling to the SM particles.

When there is no direct coupling between the dark matter and either the inflaton or SM particles,
production through gravitational interactions is always present~\cite{ema,Garny:2015sjg,Tang:2017hvq,Bernal:2018qlk,Chianese:2020yjo,Redi:2020ffc,MO,Barman:2021ugy,Ahmed:2021fvt,Bernal:2021kaj,Haque:2021mab,cmov,Haque:2022kez,Aoki:2022dzd,cmosv,cmo,moz,Ahmed:2022tfm,Barman:2022qgt,Haque:2023yra,Kaneta:2022gug,Kaneta:2023kfv,Garcia:2023qab}. These gravitational interactions include processes that produce dark matter either from gravitational scattering within the thermal bath or directly from the inflaton condensate. Both scenarios have been explored for the production of either spin 0 or spin-$\frac{1}{2}$ particles \cite{cmov} and the thermal production with a massive spin-2 mediator was considered in \cite{Bernal:2018qlk}. The dependence on spin in gravitational production is not immediately intuitive.
However, when we represent the gravitational interaction through the exchange of a massless spin-2 graviton, the relationship between spin and gravity becomes 
evident. The source of production is also important. 
In fact, inflaton scattering 
can be interpreted as the scattering of
spin-0 particles at rest in the case of quadratic potential. Using a simple helicity argument, we expect the amplitude to be proportional to the mass of a final state fermion. On the other hand, the conformal 
nature of massless spin-1 particles also leads to the 
conclusion that they cannot be produced by gravitational interactions. In conclusion, for massless final states, only scalars can be gravitationally produced by inflaton scattering.

In this work, we demonstrate that the production of spin-$\frac{3}{2}$ particles is more intricate than the previously mentioned cases.
The production of a spin-$\frac{3}{2}$ dark matter candidate from the thermal bath $\psi_\mu$ was considered in \cite{gmov}, where this particle was called the {\it raritron}. However, to produce such a raritron, it was necessary to introduce a coupling $\psi_\nu A_\mu \nu$
between the raritron, the photon, and a neutrino, implying its metastability. It is well known since the work of \cite{Velo:1969bt} that coupling a spin-$\frac{3}{2}$ particle to the electromagnetic field leads to pathologies, though these can be addressed within the supergravity framework~\cite{Das:1976ct,Deser:2001dt}.
It is important to determine whether raritrons can be produced in a generic framework solely through gravitational interactions, driven by the oscillation of the inflaton. If they are stable, this 
would correspond to the minimum amount of 
spin-$\frac{3}{2}$ fields still 
present in the Universe and contributing to the dark  sector.

The structure of this paper is as follows: In Section \ref{sec:rs}, we provide a brief review of the properties of a fundamental spin-$\frac{3}{2}$ particle. Its coupling to the graviton is discussed in Section \ref{sec:couplings}, while its production rate is explored in Section \ref{sec:prodrate}. In Section~\ref{sec:relic}, we compute the relic density generated by the oscillations of the inflaton at the end of the inflationary phase, mediated by graviton exchange. The gravitational production from the thermal bath is discussed in~\ref{sec:relth}. Finally, we apply our results to one of the best-motivated raritron models, the gravitino, in Section \ref{sec:gravdm}, and discuss a specific supergravity model in Section~\ref{sec:sugramodel}. 
In Section~\ref{sec:st-therm}, we compare our results with the standard thermal production of gravitinos in both low-scale and high-scale supersymmetric models. 
We conclude in Section \ref{sec:sum}, and provide some additional details of the calculations in Appendices~\ref{sec:tmn32}, \ref{sec:thermal}, and~\ref{sec:ratecomp}.

\section{Gravitational Spin-$\frac32$ Production}
\begin{figure}[!ht]
\centering
\includegraphics[width=3.in]{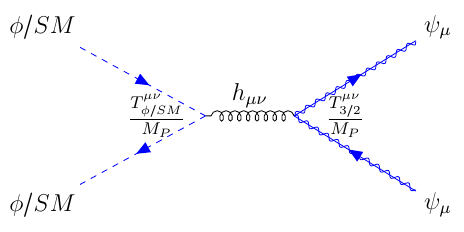}
\caption{\em \small Feynman diagram for the production of spin-$\frac32$ particles through the gravitational scattering of the inflaton condensate or the Standard Model particle bath. 
}
\label{Fig:feynman}
\end{figure}

In this section, we compute the gravitational production of a spin-$\frac{3}{2}$ particle directly from the 
inflaton condensate as well as from scatterings among  
Standard Model (SM) particles in the thermal bath.
In both 
scenarios, the interaction is mediated by the canonical 
gravitational perturbation $h_{\mu \nu}$, and 
the only distinction between the two processes is the 
source fueling the production. This perturbation arises 
when the space-time metric is expanded around the flat 
Minkowski metric, with $g_{\mu \nu} \simeq \eta_{\mu \nu} + 2 h_{\mu \nu}/M_P$. This approximation is valid 
during the reheating phase after the end of inflation. 
Importantly, such gravitational interactions are 
universal and invariably exist between the 
inflaton, the thermal bath, and the spin-$\frac{3}{2}$
raritron, as depicted in Fig.~\ref{Fig:feynman}.

\subsection{The Rarita-Schwinger field}
\label{sec:rs}

Naively, while looking at the table of known 
fundamental particles, we observe spin-0, spin-$\frac{1}{2}$, spin-1 and spin-2 fields. Naturally, one might question the absence of a spin-$\frac{3}{2}$ fundamental particle.
However, it is often claimed that 
fields with spin higher than 1 exhibit pathologies. After the 1939 paper by Fierz and Pauli
\cite{Fierz:1939ix}, where they constructed the Lagrangians for spin-$\frac{3}{2}$ and spin-2 particles,
Rarita and Schwinger proposed
a more compact formulation for spin-$\frac{3}{2}$ particles \cite{Rarita:1941mf}, leading to
the equations of motion
\beq
(i\gamma^\mu \partial_\mu - m_{3/2})\psi_\mu=0\,,~~~\gamma^\mu \psi_\mu=0
\,.
\label{Eq:eom}
\eeq
These equations can be obtained 
from the Lagrangian\footnote{For a derivation of (\ref{Eq:lagrangian32}) from Eq.~(\ref{Eq:eom}), see for example \cite{mybook}.}
\beq
{\cal L}_{3/2} \; = \; \bar \psi_\mu (i \gamma^{\mu \rho \nu} \partial_\rho + m_{3/2}\gamma^{\mu \nu})\psi_\nu\,,
\label{Eq:lagrangian32}
\eeq
with
\beq
\gamma^{\mu \nu} \; = \; \frac{1}{2}
\left[\gamma^\mu,\gamma^\nu\right]=\gamma^\mu \gamma^\nu-\eta^{\mu \nu}\,,
\eeq
and
\beq
\gamma^{\mu \nu \rho} \; = \; 
\gamma^\mu\gamma^\nu\gamma^\rho+\eta^{\mu \rho}\gamma^\nu-\eta^{\nu \rho}\gamma^\mu-\eta^{\mu \nu}\gamma^\rho\,.
\eeq
A stable $\psi_\mu$ is called a raritron and can constitute the majority of the dark matter component of the Universe.

\subsection{Gravitational Couplings}
\label{sec:couplings}

To compute the gravitational interactions between the raritron and the graviton, the space-time metric is expanded around Minkowski spacetime using $g_{\mu \nu} \simeq \eta_{\mu \nu} + \frac{2h_{\mu \nu}}{M_P}$.
The Lagrangian can then be written as~(see e.g., \cite{cmov,cmosv,hol})
\beq
\sqrt{-g}{\cal L}_{\rm int} \; = \; -\frac{1}{M_P}h_{\mu \nu}
\left(T^{\mu \nu}_{ \rm SM}+T^{\mu \nu}_\phi + T^{\mu \nu}_{\psi_\mu} \right) \, ,
\label{Eq:lagrangian}
\eeq
where SM represents Standard Model fields and $\phi$ is the inflaton. The form of the canonical stress-energy tensor $T^{\mu \nu}_i$ depends on the spin of the field, with $i = 0, \, 1/2, \, 1, \, 3/2$. For the inflaton and SM fields, we take
\bea
T^{\mu \nu}_{0} &=&
\partial^\mu S \partial^\nu S-
g^{\mu \nu}
\left[
\frac{1}{2}\partial^\alpha S \partial_\alpha S-V(S)\right] \, ,
\label{Eq:tensors}
\\
T^{\mu \nu}_{1/2} &=&
\frac{i}{4}
\left[
\bar \chi \gamma^\mu \overset{\leftrightarrow}{\partial^\nu} \chi
+\bar \chi \gamma^\nu \overset{\leftrightarrow}{\partial^\mu} \chi \right] \nonumber \\
&& -g^{\mu \nu}\left[\frac{i}{2}
\bar \chi \gamma^\alpha \overset{\leftrightarrow}{\partial_\alpha} \chi
-m_\chi \bar \chi \chi\right] \, , \\
\label{Eq:tensorf}
T_{1}^{\mu \nu} &=& \frac{1}{2} \left[ F^\mu_\alpha F^{\nu \alpha} + F^\nu_\alpha F^{\mu \alpha} - \frac{1}{2} g^{\mu \nu} F^{\alpha \beta} F_{\alpha \beta} \right] \, ,
\label{Eq:tensorv}
\eea
where $V(S)$ is the scalar potential for either the 
inflaton or the SM Higgs boson\footnote{In our calculations, we considered real scalar fields with $H$ corresponding to 4 degrees of freedom.}, with $S =  \phi,  H$, and $A\lrd_\mu B\equiv A\partial_\mu B-(\partial_\mu A)B$. Here $F_{\mu \nu}=\partial_\mu A_\nu-\partial_\nu A_\mu$
is the field strength for a vector field, $A_\mu$. 
The energy-momentum tensor for a spin-$\frac32$ Majorana field is given by\footnote{See Appendix \ref{sec:tmn32} for a detailed derivation and discussion of $T_{3/2}^{\mu\nu}$.}~\cite{tmn}
\bea
T_{3/2}^{\mu\nu}
&& =
    -\frac{i}{4}\overline\psi_\rho\gamma^{(\mu}\lrd{}^{\nu)}\psi^\rho\nonumber\\
&&+\frac{i}{2}\overline\psi{}^{(\nu}\gamma^{\mu)}\lrd_\rho\psi^\rho
    +\frac{i}{2}\overline\psi^\rho\gamma^{(\mu}\lrd_\rho\psi^{\nu)}\, ,
    \label{tmn32}
\eea
where parentheses surrounding indices indicate symmetrization, defined by $A^{(\mu}B^{\nu)}\equiv (A^\mu B^\nu + A^\nu B^\mu)/2$. For a Dirac spin-3/2 field instead, the right-hand side of Eq.~(\ref{tmn32}) should be multiplied by a factor of 2.

The gravitational scattering amplitudes related to the production rate of the processes
\beq
\phi/{\rm{SM}}^i(p_1)+\phi/{\rm{SM}}^i(p_2) \rightarrow \psi(p_3)+ \psi(p_4)
\label{Eq:process}
\eeq
can be parametrized by
\begin{equation}
\mathcal{M}^{i \frac32} \propto M_{\mu \nu}^\frac32 \Pi^{\mu \nu \rho \sigma} M_{\rho \sigma}^i \;, 
\label{scamp}
\end{equation}
where $i=0,1/2,1$ denotes the spin of the initial state involved in the scattering process. Note that we are summing over all polarizations, justifying the absence of Lorentz indices in Eq.~(\ref{Eq:process}) for the raritron. 
Here, $\Pi^{\mu \nu \rho \sigma}$ is the graviton propagator for the canonical field $h_{\mu \nu}$ with momentum $k = p_1+p_2$,
\begin{equation}
 \Pi^{\mu\nu\rho\sigma}(k) = \frac{\eta^{\mu\rho}\eta^{\nu\sigma}
 +\eta^{\mu\sigma}\eta^{\nu\rho} 
 - \eta^{\mu\nu}\eta^{\rho\sigma} }{2k^2} \, .
\end{equation} 
 The partial amplitudes, $M_{\mu \nu}^i$, can be expressed by \cite{cmov}
\bea 
\label{partamp1}
M_{\mu \nu}^{0} &=& \frac{1}{2}\left[p_{1\mu} p_{2\nu} + p_{1\nu} p_{2\mu} - \eta_{\mu \nu}p_1\cdot p_2 - \eta_{\mu \nu} V''(S)\right] \,, \\ 
M_{\mu \nu}^\frac12 &=&  \frac{1}{4} {\bar v}(p_2) \left[ \gamma_\mu (p_1-p_2)_\nu + \gamma_\nu (p_1-p_2)_\mu \right] u(p_1) \, , \\
M_{\mu \nu}^{1} &=& \frac{1}{2} \bigg[ \epsilon_{2}^{*} \cdot \epsilon_{1}\left(p_{1 \mu} p_{2 \nu}+p_{1 \nu} p_{2 \mu}\right)
\nonumber\\
\label{partamp2}
&-&\epsilon_{2}^{*} \cdot p_{1}\left(p_{2 \mu} \epsilon_{1 \nu}+\epsilon_{1 \mu} p_{2 \nu}\right) - \epsilon_{1} \cdot p_{2}\left(p_{1 \nu} \epsilon_{2 \mu}^{*}+p_{1 \mu} \epsilon_{2 \nu}^{*}\right)
\nonumber\\
\label{partamp3}
&+&p_{1} \cdot p_{2}\left(\epsilon_{1 \mu} \epsilon_{2 \nu}^{*}+\epsilon_{1 \nu} \epsilon_{2 \mu}^{*}\right)  \nonumber \\
&+&\eta_{\mu \nu}\left(\epsilon_{2}^{*} \cdot p_{1} \epsilon_{1} \cdot p_{2}-p_{1} \cdot p_{2} \, \epsilon_{2}^{*} \cdot \epsilon_{1}\right) \bigg]  \, ,
\eea
where the masses of the SM fermions and vector fields have been neglected. 
The partial amplitude for the Majorana spin-$\frac32$ field $\psi_{\mu}$
is given by 
\bea 
M_{\mu \nu}^\frac32 & = & \frac{1}{4} \left[  {\bar v}^\alpha(p_4) \gamma_{(\mu} (p_3-p_4)_{\nu)} u_\alpha(p_3) \right. \nonumber \\
&-& 2\left. {\bar v}^\alpha(p_4) \gamma_{(\mu} (p_3-p_4)_\alpha u_{\nu)}(p_3) \right.\nonumber \\& -& 2\left. {\bar v}_{(\nu}(p_4) \gamma_{\mu)} (p_3-p_4)_\alpha u^{\alpha}(p_3) \right] \, ,
\label{amp32}
\eea
where we defined $\psi_\mu(p)=u_\mu(p)e^{-ipx}$.
In this section, we do not rely on any specific model of inflation and keep our discussion as general as possible. Any model satisfying the constraints on the slow-roll parameters as imposed by \textit{Planck} data~\cite{Planck} will suffice, provided there is a well-defined minimum and the potential can be expanded as $V(\phi) \simeq \lambda \phi^k/M_P^{k-4}$
around this minimum. For example, both the Starobinsky model \cite{Staro} and $\alpha$-attractor type models \cite{Kallosh:2013hoa} are sufficient.

We consider two distinct processes illustrated by the Feynman diagram in Fig.~\ref{Fig:feynman}: 

i) The production of raritrons from the inflaton $\phi + \phi \rightarrow \psi + \psi$. 
In the case of a quadratic potential, 
the inflaton behaves like a massive particle at rest,\footnote{Up to
some symmetry factors, see \cite{MO,cmov}.} with four-momentum $p_{1,2}$. Its partial amplitude is then directly given by Eq.~(\ref{partamp1}). 
However, for a generic potential $V(\phi)$,
we need to use the zero mode of the inflaton condensate that is valid for any arbitrary minimum 
(see below and \cite{cmosv} for a detailed discussion). 

ii) The production from the thermal background, $\textrm{SM} + \textrm{SM}  \rightarrow \psi + \psi$, which uses Eqs.~(\ref{Eq:tensors}-\ref{Eq:tensorv}) for SM particles on the right-hand side of Eq.~(\ref{scamp}).  In the following subsection, we compute  the full scattering amplitudes for both channels and determine the gravitational production rate of the raritron.

\subsection{Gravitational Production from the Inflaton Condensate}
\label{sec:prodrate}

We begin by examining the gravitational production of the raritron from the inflaton condensate. Although particle production occurs throughout the reheating process, the dominant source of energy density emerges at the onset of oscillations after inflation, when the oscillation amplitude peaks. Notably, despite gravitational production process being Planck-suppressed, the inflaton condensate scattering continues to be a substantial source of particle production, particularly at the beginning of the reheating process, when its energy density is very large.

\subsubsection{Quadratic potential minimum}

We consider the case with a quadratic minimum first, with $V(\phi) \simeq \frac{1}{2} m_{\phi}^2 \phi^2$. In this case, the rate computation is straightforward. We evaluate the square of the matrix element in Eq.~(\ref{scamp}) using $M^0_{\rho\sigma}$ for the inflaton incoming state. We assume for the inflaton condensate that the incoming inflaton momentum vanishes, $p_{1, 2} = 0$, and compute $|\mathcal{M}^{0\frac32}|^2$ using the spinor sums \cite{Haberzettl:1998rw,Ding:2012sm}
\begin{align}
P_{ab} &\; = \; \sum_{s=-3/2}^{+3/2} u_a(\mathbf{p},s) \bar{u}_b(\mathbf{p},s) \; = \; \left(\slashed{p}+m_{3/2} \right) \times ~\nonumber\\ 
  &\left(\eta_{ab} - \frac13 \gamma_a \gamma_b - \frac23 \frac{p_a p_b}{m_{3/2}^2} + \frac{p_a \gamma_b - p_b \gamma_a}{3m_{3/2}}\right) \, ,
\label{upol}
\end{align}
and 
\begin{align}
Q_{ab} &\; = \;  \sum_{s=-3/2}^{+3/2} v_a(\mathbf{p},s) \bar{v}_b(\mathbf{p},s) \; = \; \left(\slashed{p}-m_{3/2} \right) \times ~\nonumber\\ 
  &\left(\eta_{ab} - \frac13 \gamma_a \gamma_b - \frac23 \frac{p_a p_b}{m_{3/2}^2} - \frac{p_a \gamma_b - p_b \gamma_a}{3m_{3/2}}\right) \, .
\label{upol}
\end{align}
Using the above expressions, we find that the total matrix element squared is given by Eq.~(\ref{fullM2}) shown in Appendix \ref{sec:thermal}.
For the inflaton condensate, this expression can be simplified significantly by writing 
$t = m_{3/2}^2 - m_\phi^2$ and $s = 4 m_\phi^2$, and the matrix element squared~(\ref{fullM2}) becomes
\begin{align}
\label{matelsq1}
&|\overline{\mathcal{M}}|^2  \; = \; \frac{m_{\phi}^4s}{18M_P^4m_{3/2}^2}\left(1 - \frac{4m_{3/2}^2}{s} \right)\left(1 - \frac{6m_{3/2}^2}{s} + \frac{18m_{3/2}^4}{s^2} \right)\nonumber \\
& =  \frac{2}{9} \frac{m_{\phi}^6}{m_{3/2}^2 M_P^4} \left(1 - \frac{m_{3/2}^2}{m_{\phi}^2} \right) \left(1 - \frac{3}{2} \frac{m_{3/2}^2}{m_{\phi}^2} + \frac{9}{8} \frac{m_{3/2}^4}{m_{\phi}^4} \right)\, .
\end{align}
The production rate, $R^{\phi^k}$, for a quadratic minimum with $k = 2$, can be written as \cite{mybook}
\beq
R^{\phi^2} \; = \; 
n_\phi^2\langle \sigma v \rangle =
\frac{\rho_\phi^2}{m_\phi^2}\frac{|\overline{\mathcal{M}}|^2}{32\pi m_\phi^2} \frac{p_3}{m_\phi}  \, ,
\label{rate2}
\eeq
where $p_3 = \sqrt{m_\phi^2 - m_{3/2}^2}$, and if we use the matrix element squared~(\ref{matelsq1}), we find
\beq
R^{\phi^2} \; =  \frac{2 \times \rho_\phi^2}{288 \pi M_P^4}  \left(\tau^{-1} - \frac{3}{2} + \frac{9}{8} \tau \right) \left(1 - \tau\right)^{3/2} \, ,
\label{inflrate}
\eeq
with $\tau = m_{3/2}^2/m_\phi^2$.  The factor of 2 explicitly accounts for the fact that 2 raritrons are
produced by annihilation.\footnote{We note that in Appendix \ref{sec:thermal}, we provide the amplitude in the case of scalar scattering, which yields a rate which is larger by a factor of {$2$} compared 
to inflaton scattering when considering a condensate $\phi$.}

Before extending our result to 
a more general potential $V(\phi)$, we would like to make a few comments regarding Eq.~(\ref{inflrate}). The massive raritron
could have been considered, naively, as a Clebsch-Gordan decomposition of a massive spin-1 boson and a spin-$\frac12$ fermion, which is manifestly not the case when we look at the limit $\tau \rightarrow 0$
of Eq.~(\ref{inflrate}). Indeed, we would expect no production of massless fermions, due to helicity conservation \cite{MO}, and no divergences are expected for the production
of a massless vector field \cite{Barman:2021ugy}. This reflects the inherent pathology of theories with spin $>1$, implying that we should treat with care the
unitarity constraints when we analyze the bounds on $m_{3/2}$.

\subsubsection{General potentials}

Gravitational particle production from the inflaton condensate naturally depends on the shape of the potential. We extend our discussion and consider a more general potential which about its minimum is of the form
\begin{equation}
    \label{eq:potosc}
    V(\phi) \; = \; \lambda \frac{\phi^k}{M_P^{k-4}} \, , \qquad \phi \ll M_P \, .
\end{equation}
We parameterize the time-dependent oscillating inflaton field as
\begin{equation}
    \label{Eq:oscillation}
    \phi(t) \; = \; \phi_0(t) \cdot \mathcal{P}(t) \, , 
\end{equation}
where $\phi_0(t)$ is the time-dependent envelope that includes the effects of redshift and $\mathcal{P}(t)$ describes the periodicity of the oscillation. Then for a potential of the form~(\ref{eq:potosc}), we can write $V(\phi) = V(\phi_0) \cdot \mathcal{P}(t)^k$ and expand the potential energy in terms of its Fourier modes \cite{Ichikawa:2008ne,Kainulainen:2016vzv,GKMO2}
\beq
V(\phi)=V(\phi_0)\sum_{n=-\infty}^{\infty} {\cal P}_{k,n} e^{-in \omega t}
=\langle \rho_\phi \rangle \sum_{n = -\infty}^{\infty} {\cal P}_{k, n} e^{-in \omega t} \, ,
\eeq
where $\omega$ is the frequency of oscillation of $\phi$, given by \cite{GKMO2}
\beq
\label{eq:angfrequency}
\omega=m_\phi \sqrt{\frac{\pi k}{2(k-1)}}
\frac{\Gamma(\frac{1}{2}+\frac{1}{k})}{\Gamma(\frac{1}{k})} \, ,
\eeq
with $m_\phi^2 = \partial^2 V/\partial \phi^2|_{\phi_0}$, 
\begin{equation}
    \mathcal{P}(t)^k \; = \; \sum_{n = -\infty}^{\infty} \mathcal{P}_{k, n} e^{- i n \omega t} \,,
\end{equation}
and $\langle \rho_\phi \rangle$ is the mean energy density averaged over the oscillations.

To compute the inflaton condensate scattering rate, we follow the treatment presented in Appendix~\ref{sec:ratecomp}. We find that the raritron production rate is given by
\begin{equation}
    \label{eq:rategenk}
    R^{\phi^k} \; = \; \frac{2 \times \rho_{\phi}^2}{72 \pi M_P^4}  \Sigma_{3/2}^k \, ,
\end{equation}
where
\begin{align}
    \label{sigma32}
    \Sigma_{3/2}^k \; = \; \sum_{ n = 1}^{+\infty} |\mathcal{P}_{k, n}|^2 \frac{E_n^2}{m_{3/2}^2}\left(1-6 \frac{m_{3/2}^2}{E_n^2} + 18 \frac{m_{3/2}^4}{E_n^4} \right) \times~\nonumber \\
    \times\left[1 - \frac{4m_{3/2}^2}{E_n^2} \right]^{3/2} \, .
\end{align}
Here the superscript $k$ corresponds to the type of potential minimum $V(\phi) \sim \phi^k \sim \mathcal{P}^k$,  $E_n = n \omega$ is the energy of the $n$-th mode of the inflaton oscillation, and $m_{3/2}$ is the produced raritron mass.
In the quadratic case, where $\omega = m_\phi$ (see Eq.~(\ref{eq:angfrequency})) and
${\cal P}(t)^2=\cos^2 (m_\phi t)=\frac{1}{2}+\frac{1}{4}(e^{-2 m_\phi t}+e^{2 m_\phi t})$,
since 
$\sum|\mathcal{P}_{2, n}|^2 =|{\cal P}_{2,2}|^2 = \frac{1}{16}$, only the second mode in the Fourier expansion contributes to the sum.
Taking $E_2 = 2 m_{\phi}$, 
we find that the rate~(\ref{eq:rategenk}) reduces to Eq.~(\ref{rate2}).

We also consider separately the production of the $\pm \frac12$ and $\pm \frac32$ helicity components. One can express the spin-$\frac32$ polarization vector as a direct product of spin-1 and spin-$\frac12$ polarization vectors. We introduce the following spin-$\frac32$ Clebsch-Gordan decomposition for the spinor\footnote{As a side comment, from this decomposition, one can also derive the spinor-helicity formalism for massive spin-3/2 fields, and compute helicity amplitudes. See for example \cite{Diaz-Cruz:2016abv}.}
\bea
u^{\mu}_{\pm 3/2}(p) & = & \epsilon^{\mu}_{\pm}(p) u_{\pm 1/2}(p) \, , 
\label{Eq:p32}\\
u^{\mu}_{\pm 1/2}(p)  & = & \sqrt{\frac23} \epsilon_0^{\mu}(p)u_{\pm 1/2}(p)~\nonumber \\
 &&+\frac{1}{\sqrt{3}}  \epsilon^{\mu}_{\pm}(p) u_{\mp 1/2}(p) \, , 
 \label{Eq:p12}\\
v^{\mu}_{\pm 3/2}(p) & = & \epsilon^{\mu*}_{\pm}(p) v_{\pm 1/2}(p) \, , 
\label{Eq:m32}\\
v^{\mu}_{\pm 1/2}(p)  & = & \sqrt{\frac23} \epsilon_0^{\mu*}(p)v_{\pm 1/2}(p)~\nonumber \\
&& +\frac{1}{\sqrt{3}} \epsilon^{\mu*}_{\pm}(p) v_{\mp 1/2}(p) \, . 
\label{decomp}
\eea

We find that the raritron production rate~(\ref{eq:rategenk}) can be decomposed as
\begin{equation}
    \label{eq:rategenkd}
    R^{\phi^k} \; = \; \frac{2 \times \rho_{\phi}^2}{72 \pi M_P^4}  \left(\Sigma_{3/2, 3/2}^k + \Sigma_{3/2, 1/2}^k \right)\, ,
\end{equation}
where the transverse spin $\pm \frac{3}{2}$ contribution is given by
\begin{align}
    \Sigma_{3/2, 3/2}^k \; = \; \sum_{n = 1}^{+\infty} |\mathcal{P}_{k, n}|^2 \frac{E_n^2}{m_{3/2}^2} \times \left(9 \frac{m_{3/2}^4}{E_n^4} \right) \times~\nonumber \\
    \times\left[1 - \frac{4m_{3/2}^2}{E_n^2} \right]^{3/2} \, ,
    \label{trans}
\end{align}
and the longitudinal spin $\pm \frac{1}{2}$ contribution is
\begin{align}
     \Sigma_{3/2, 1/2}^k \; = \; \sum_{n = 1}^{+\infty} |\mathcal{P}_{k, n}|^2 \frac{E_n^2}{m_{3/2}^2} \times \left(1 - 3\frac{m_{3/2}^2}{E_n^2} \right)^2 \times~\nonumber \\
    \times\left[1 - \frac{4m_{3/2}^2}{E_n^2} \right]^{3/2} \, .
    \label{long}
\end{align}
We note that the sum of transverse and longitudinal components satisfy the expression~(\ref{sigma32}), with $\Sigma_{3/2}^k = \Sigma_{3/2, 3/2}^k + \Sigma_{3/2, 1/2}^k$. 

Returning to the pathology of the limit $m_{3/2}\rightarrow 0$
($\tau \rightarrow 0$) discussed above, we observe that the transverse components $\pm \frac32$ are not produced 
for $m_{3/2}=0$. 
These components correspond to a direct composition between a spin-$\frac12$ fermion and the transverse components of a spin-1, as we can see in Eqs.~(\ref{Eq:p32}) and (\ref{Eq:m32}). As these transverse components are not gravitationally produced for massless particles \cite{MO}, it stands to reason that their production rate vanishes in the massless limit
for a spin-$\frac32$ particle as well. 
In other words, the transverse modes are expected to be highly suppressed for light raritron, relative to the 
longitudinal mode which is enhanced and could be considered as the {\it goldstino} in a gauged framework.

\begin{figure}[!ht]
  \centering
\includegraphics[width=0.45\textwidth]{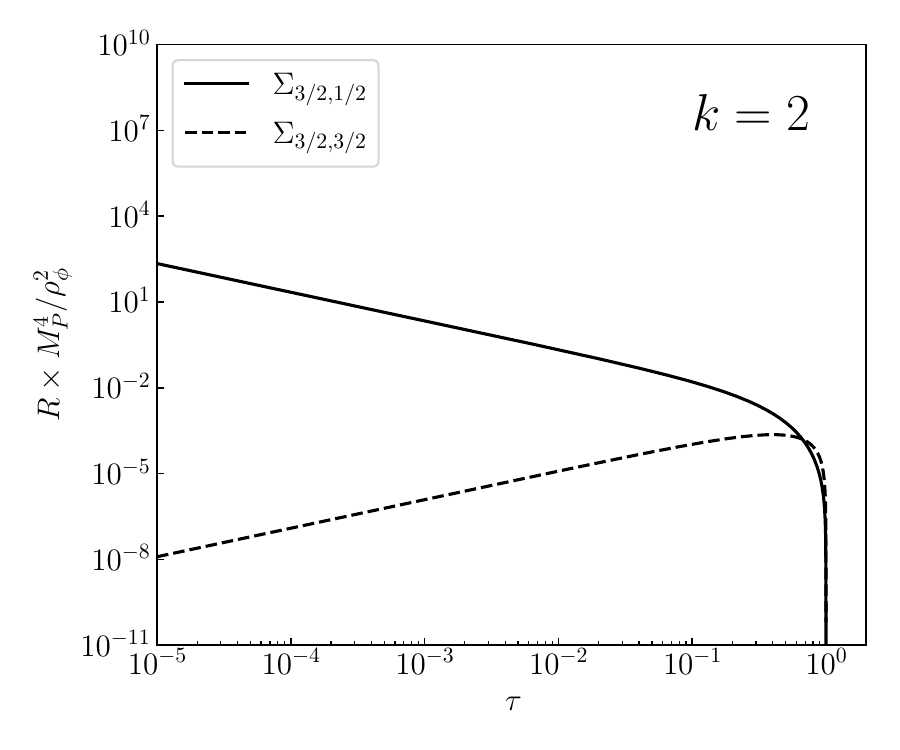}
  \caption{\em \small Longitudinal and transverse raritron production rates for $k=2$ in the units of $R \times M_P^4/\rho_{\phi}^2$ as a function of $\tau = m_{3/2}^2/m_{\phi}^2$. As can be seen from the figure, the raritron production is completely dominated by the longitudinal component, which contains a factor $\tau^{-1}$.}
  \label{Fig:rates}
\end{figure}

In Fig.~\ref{Fig:rates},
we plot separately the longitudinal and transverse components  for $k = 2$.
We clearly see the effect we have just described: the transverse mode is always produced in 
negligible quantities compared with the longitudinal mode, except in the limit where the 
mass of the raritron is of the order of the fundamental mode $m_{3/2}\simeq m_\phi$. The 
slopes for masses $m_{3/2}\lesssim m_\phi$,
keeping only the first Fourier mode as an approximation, gives 
$R_{\pm 3/2}\propto m_{3/2}^2$ and $R_{\pm 1/2}\propto {m_{3/2}^{-2}}$ and do not depend on $k$.
The absolute value of the rates  depends on the Fourier coefficients ${\cal P}_{n,k}$
which themselves become very similar for large values of $k$, hence we do not expect big differences for larger values of $k$.

\subsection{Relic abundance calculation}
\label{sec:relic}

Given the production rate,
we next compute the abundance of raritrons from the Boltzmann equation,
\beq
\frac{dn}{dt} + 3Hn = R^{\phi^k} \,,
\label{Eq:boltzmann1}
\eeq
where $H=\frac{\dot a}{a}$ is the Hubble parameter. 
It is convenient to rewrite the Boltzmann equation in terms of the scale factor, 
\beq
\frac{dY}{da}=\frac{a^2R^{\phi^k}}{H} \, ,
\label{Eq:boltzmann2}
\eeq
where $Y \equiv a^3 n$. 
To integrate this expression we need to include the dependence $H(a)$
with
\beq
H(a)=\frac{\rho^\frac{1}{2}_\phi(a)}{\sqrt{3}M_P} \, .
\eeq
The conservation of energy for the inflaton field imposes
\bea
\frac{d \rho_\phi}{dt} + 3(1+w)H \rho_\phi &= & Ha\left[\frac{d \rho_\phi}{da}+3(1+w)\frac{\rho_\phi}{a}\right] \nonumber \\ &=&(1+w)\Gamma_\phi \rho_\phi \, ,
\eea
whose solution is, for $\Gamma_\phi \ll H$
\beq
\rho_\phi(a)=\rho_{\rm end}\left(\frac{a_{\rm end}}{a}\right)^\frac{6k}{k+2}=\rho_{\rm RH}\left(\frac{a_{\rm RH}}{a}\right)^{\frac{6k}{k+2}}\,.
\label{rhophia}
\eeq
In these expressions, $a_{\rm end}$ is the value of the scale factor when accelerated expansion (inflation) ends, $\rho_{\rm end} = \rho_\phi(a_{\rm end})$, $a_{\rm RH}$ is the scale factor when  $\rho_R(a_{\rm RH})=\rho_\phi(a_{\rm RH})$, defining the moment of reheating.
The Boltzmann equation~(\ref{Eq:boltzmann2}) then becomes
\beq
\frac{dY}{da}=\frac{\sqrt{3}M_P}{\sqrt{\rho_{\rm RH}}}a^2\left(\frac{a}{a_{\rm RH}}\right)^{\frac{3k}{k+2}}R^{\phi^k}(a) \, .
\label{Eq:boltzmann4}
\eeq

Restricting our attention to the case $k = 2$,
we have $\rho_\phi \sim a^{-3}$, $\rho_R \sim T^4 \sim a^{-3/2}$, with $m_\phi^2 = 2 \lambda M_P^2$. The Boltzmann equation becomes
\beq
\frac{dY}{da}=\frac{\sqrt{3}M_P}{\sqrt{\rho_{\rm RH}}}a^2\left(\frac{a}{a_{\rm RH}}\right)^{\frac32}R^{\phi^2}(a) \, ,
\label{Eq:boltzmann5}
\eeq
where $R^{\phi^2}(a)$ is given by Eq.~(\ref{inflrate}). 
Eq.~(\ref{Eq:boltzmann5}) is easily integrated  to give 
\begin{eqnarray}
n(a_{\rm RH}) & = & \frac{1}{72 \sqrt{3} \pi  M_P} \left(\frac{\rho_{\rm end}}{M_P^4} \right)^\frac12 \alpha T_{\rm RH}^4 \nonumber \\
&& \times \left(\tau^{-1} -\frac32 +\frac98 \tau \right) \left(1 - \tau\right)^{3/2} \, ,
\label{n32tot}
\end{eqnarray}
where we assumed that $a_{\rm RH} \gg a_{\rm end}$ and 
$\alpha$ is defined by
\begin{equation}
    \rho_R \; = \; \frac{g_T \pi^2}{30} T^4 \equiv \alpha T^4 \, .
\end{equation}
Using \cite{mybook}
\beq
\label{eq:relicabund1}
\Omega h^2 \simeq 1.6\times 10^8\frac{g_0}{g_{\rm RH}}\frac{n(\trh)}{\trh^3}\frac{m_{3/2}}{1~{\rm GeV}} \, ,
\eeq
we then obtain
\begin{eqnarray}
\Omega h^2 & \simeq & 3 \times 10^{9} \left( \frac{T_{\rm RH}}{10^{10} {\rm GeV}} \right) \left( \frac{\rho_{\rm end}}{(5.2 \times 10^{15} {\rm GeV})^4} \right)^\frac12 \nonumber \\ & & \left( \frac{m_\phi}{1.7 \times 10^{13} {\rm GeV}} \right)^2 \left( \frac{{\rm EeV}}{m_{3/2}} \right) \, ,
\label{oh2tot}
\end{eqnarray}
where we take $g_0=43/11$ and $g_{\rm RH}=427/4$, and assume $m_{3/2} \ll m_\phi$, and values of $m_\phi$ and $\rho_{\rm end}$ are 
normalized for an $\alpha$-attractor model of inflation with $k=2$, though there 
is some additional dependence on $T_{\rm RH}$ for these quantities \cite{GKMO1,cmov,egnov}.

\begin{figure}[!ht]
\includegraphics[width=0.5\textwidth]{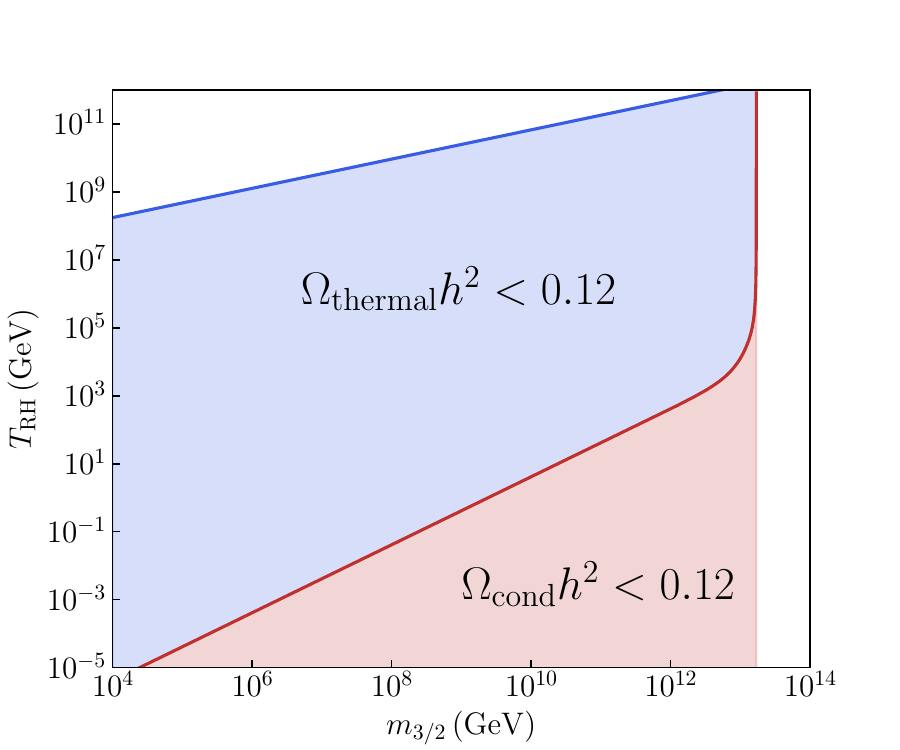}
\caption{\em \small Contours of $\Omega_{\rm cond} h^2 = 0.12$ (red) and $\Omega_{\rm thermal} h^2 = 0.12$ (blue) in the $(m_{3/2}, \trh)$ plane.}
\label{Fig:thermalcond}
\end{figure}

As one can see, the gravitational production of raritrons is extremely efficient, much more efficient
than the production of scalars, spin-$\frac12$ fermions, or vectors \cite{MO}.
As a consequence,  stable raritrons are only possible if either $\trh$ is quite low (of order the weak scale or below) or $m_{3/2} \simeq m_\phi$. This can be seen in Fig.~\ref{Fig:thermalcond} where the red curve shows the values of $m_{3/2}$ and $\trh$ such that $\Omega h^2 = 0.12$ from Eq.~(\ref{oh2tot}). To remain consistent with big bang nucleosynthesis, $\trh \gtrsim 2$ MeV,  which implies that raritron dark matter is heavier than $\gtrsim 6$ PeV. This {\it unavoidable} (because it is produced gravitationally) minimal mass for the raritron is one of the main results of our work.

Using Eqs.~(\ref{eq:rategenkd})-(\ref{long}), it is possible to separate out the contributions of the transverse and longitudinal contributions to $\Omega_\frac32$. 
For the transverse contribution for $k=2$ we have
\beq
n_\frac32(a_{\rm RH}) = \frac{1}{128 \sqrt{3} \pi  M_P} \left(\frac{\rho_{\rm end}}{M_P^4} \right)^\frac12 \alpha T_{\rm RH}^4 \tau(1-\tau)^\frac32 \, ,
\label{ntotgrav}
\eeq
and
\begin{eqnarray}
\Omega_\frac32 h^2 & \simeq & 2 \times 10^{-8} \left( \frac{T_{\rm RH}}{10^{10}~ {\rm GeV}} \right) \left( \frac{\rho_{\rm end}}{(5.2 \times 10^{15} {\rm GeV})^4} \right)^\frac12 \nonumber \\ & & \left( \frac{1.7 \times 10^{13} {\rm GeV}}{m_\phi} \right)^2 \left( \frac{m_{3/2}}{{\rm EeV}} \right)^3 \, .
\label{oh2totgrav}
\end{eqnarray}
As expected and discussed previously, the gravitational production of the transverse mode is completely negligible.

Similarly, we can compute the longitudinal contribution
\begin{eqnarray}
n_{\frac{1}{2}}(a_{\rm RH}) & = & \frac{1}{72 \pi \sqrt{3} M_P} \left(\frac{\rho_{\rm end}}{M_P^4} \right)^\frac12 \alpha T_{\rm RH}^4 \nonumber \\
&& \times \left(\tau^{-1} -\frac32 +\frac{9}{16} \tau \right) \left(1 - \tau\right)^{3/2} \, ,
\end{eqnarray}
which for
$m_{3/2} \ll m_\phi$, gives the result in Eq.~(\ref{oh2tot}) for $\Omega_\frac12 h^2$ since the production of raritrons is completely dominated by the longitudinal component which carries the factor of $\tau^{-1}$.

At this point it is important to note that for ``low'' values of $\tau$, we may run into a problem with unitarity. The amplitude in Eq.~(\ref{matelsq1})
becomes of order unity when $m_{3/2} \lesssim 1$ TeV. However, 
raritron scattering $\psi_\mu\psi_\mu\to h_{\mu\nu}\to\psi_\mu\psi_\mu$ is further enhanced, and we estimate that its amplitude scales as 
 $|\mathcal{M}|^2 \propto m_\phi^4/(M_P^4 \tau^4)$, which would exceed unity when $m_{3/2} \lesssim 40$ EeV!\footnote{We consider here a non-supersymmetric theory where the spin-$\frac32$ Lagrangian is given by \eqref{Eq:lagrangian32}. When supersymmetry is introduced, an additional contribution to raritron scattering arises from the four-Fermi coupling, which cancels the most divergent term in the amplitude, leading to $|\mathcal{M}|^2 \propto m_\phi^4/(M_P^4 \tau^2)$ \cite{Antoniadis:2022jjy}. In this case, unitarity is violated when $m_{3/2} \lesssim 0.1$ EeV.}
This would allow reheating temperatures $\trh \gtrsim 10$ GeV. 

In addition to problems with unitarity, low mass raritrons are produced with low and potentially vanishing sound speeds \cite{Kolb:2021xfn}. The gravitational production of raritrons was calculated by solving the mode function with Bunch-Davies initial conditions.\footnote{The production of scalars using mode functions was recently compared \cite{Kaneta:2022gug} to the perturbative production \cite{MO,cmov}, and the two were found to be comparable so long as the low momentum modes are suppressed, e.g., due to self-interactions or the conformal coupling to the scalar curvature. } Indeed, in \cite{Kolb:2021xfn}, it was argued that sound speed vanishes whenever $m_{3/2} \lesssim 0.39 H(a_{\rm end})$. In the models considered here, $H(a_{\rm end}) \simeq 0.4 m_\phi$ and thus the sound speed vanishes when $\tau < 0.02$ or when $m_{3/2} < 2.5 \times 10^{12}$ GeV. As a consequence, the spectral densities continue to rise as the cube of the wavenumber without bound, leading to potentially ``catastrophic'' production. In contrast, for larger raritron masses, the spectral density turns over and peaks for wavenumbers of order a few $H(a_{\rm end})$. However it is difficult to compare this result with our own given in Eqs.~(\ref{n32tot}) and (\ref{oh2tot}). While we have neglected the effects of curvature in $T_{\mu\nu}$, we estimate that this may change our result by factors of order unity, while still neglecting the effects for vanishing sound speed. To determine the total number density or the relic density of raritrons from the approach using mode functions, one must specify an ultraviolet cutoff, $\Lambda$, and integrate the spectral density up to that cutoff. As the density scales as the $\Lambda^3$, it is not immediately clear that a vanishing sound speed leads to an over-density of raritrons.

Finally, we can also generalize this result to cases when $k \neq 2$. We find that the number density can be expressed as
\begin{eqnarray}
    n(a_{\rm RH}) \; = \; \frac{(k+2)}{(k-1)}\frac{\rho_{\rm RH}^{3/2}}{72\sqrt{3} \pi M_P^{3}} \left(\frac{\rho_{\rm end}}{\rho_{\rm RH}}\right)^{1-\frac{1}{k}}\Sigma_{3/2}^k \, ,
\end{eqnarray} 
and the dark matter abundance becomes
\begin{align}
&\Omega h^2\simeq 2.2 \times 10^{5}\frac{(k+2)}{(k-1)} \left(\frac{\rho_{\rm end}}{\rho_{\rm RH}} \right)^{1-\frac{1}{k}} \frac{\rho_{\rm RH}^{3/4}}{M_P^3} \frac{m_{3/2}}{1~{\rm GeV}} \Sigma_{3/2}^k  \, .
\label{oh2totgenk1}
\end{align}
As expected, for $k=2$ this expression reduces to Eq.~(\ref{oh2tot}).

\subsection{Gravitational Production of Raritrons from the Thermal Bath}
\label{sec:relth}

The production of raritron dark matter from the thermal bath is also possible. The scattering of SM particles includes the Higgs scalars, gauge bosons, and fermions in the initial state. Since the initial particle momenta $p_1$ and $p_2$ are large (of order $m_\phi$) at the beginning of reheating and dominate over electroweak scale quantities,
we assume that the initial particle states are massless.

For the Higgs initial state we can use Eq.~(\ref{fullM2}) with the association $\phi \to h$ and set $m_\phi = 0$ (i.e., we neglect the Higgs mass, and all SM masses relative to the reheating temperature in the thermal bath). 
In this case, Eq.~(\ref{fullM2}) reduces to
\begin{equation}
\begin{aligned}
&|{\cal M}^0|^2=
\frac{1}{72 m_{3/2}^4 M_P^4 s^2}\left[-s^2 t(s+t)(s+2 t)^2-72 m_{3/2}^{12}\right. \\
& +24 m_{3/2}^{10}(7 s+12 t)-2 m_{3/2}^8\left(47 s^2+264 s t+216 t^2\right)\\
& +m_{3/2}^6\left(-2 s^3+244 s^2 t+576 s t^2+288 t^3\right)\\
& +m_{3/2}^4\left(s^4-34 s^3 t-210 s^2 t^2-240 s t^3-72 t^4\right)\\&\left.+m_{3/2}^2 s\left(s^4+6 s^3 t+44 s^2 t^2+64 s t^3+24 t^4\right)\right] \, . 
\end{aligned}
\end{equation}
In addition to the production of raritrons from a Higgs initial state, other SM particles in thermal bath will also lead to raritron production.  The amplitudes for massless fermion and gauge boson initial states are given in Eqs.~(\ref{M1232}) and (\ref{M132}) respectively.

The dark matter production rate $R(T)$ for the SM+SM $\rightarrow$ $\psi + \psi$ process with amplitude $\cal{M}$ is given by\footnote{We note that we include the symmetry factors associated with identical initial and final states in the squared amplitude, $|\overline{\mathcal{M}}|^2$.}
\cite{Benakli:2017whb,grav2,mybook}
\begin{eqnarray}
R(T) & = & \frac{2}{1024 \pi^6}\times\int f_1 f_2 E_1 \diff E_1 E_2 \diff E_2 \diff \cos \theta_{12} \nonumber \\
& & \times \int |\overline{\cal M}|^2 \diff \Omega_{13} \, ,
\label{thrate}
\end{eqnarray}
where we assumed that $s \gg 4m_{3/2}^2$, and the factor of two accounts for two raritrons produced per scattering, $E_i$ denotes the energy of particle $i=1,2,3,4$.
$\theta_{13}$ and $\theta_{12}$ are the angles formed by momenta ${\bf{p}}_{1,3}$ (in the center-of-mass frame) and ${\bf{p}}_{1,2}$ (in the laboratory frame), respectively. The infinitesimal solid angle in the above integral is then $\diff\Omega_{13}=2\pi \diff\cos\theta_{13}$.  In addition, 
\beq
f_{i}= \frac{1}{e^{E_i/T}\pm 1} \, ,
\eeq
represents the assumed thermal distributions of the incoming SM particles. 

The total amplitude squared for the gravitational scattering process SM+SM $\rightarrow$ $\psi + \psi$ is given by a sum of the three amplitudes associated with three different SM initial state spins,
\begin{equation}
    \label{Eq:ampscat}
    |\overline{\mathcal{M}}|^{2}=  4 |\overline{\mathcal{M}}^{0}|^{2}+45|\overline{\mathcal{M}}^{1 / 2}|^{2}+12|\overline{\mathcal{M}}^{1}|^{2} \, .
\end{equation}
Using this amplitude and performing the thermal integration in Eq.~(\ref{thrate}), we find that the 
raritron production rate can be parameterized by
\begin{equation}
\begin{aligned}
R^T_\frac32=&R_\frac32(T)= \beta_1\frac{T^{12}}{m_{3/2}^4 M_P^4} + \beta_2\frac{T^{10}}{m_{3/2}^2 M_P^4} \\
&+\beta_3 \frac{T^{8}}{M_P^4} + \beta_4 \frac{m_{3/2}^2 T^6}{M_P^4} + \beta_5 \frac{m_{3/2}^4 T^4}{M_P^4}
\, ,
\label{Eq:ratethermal}
\end{aligned}
\end{equation}
where the numerical coefficients together with the details of the computation are given in Appendix~\ref{sec:thermal}.

The gravitational scattering within the 
thermal plasma produces the raritrons. We focus on the case $k=2$ and show that the thermal 
production rate is strongly 
sub-dominant compared to the 
production from the inflaton condensate. Following the same steps 
as in the previous subsection, we 
replace the rate in Eq.~(\ref{Eq:boltzmann4}) by the thermal 
raritron production rate~(\ref{Eq:ratethermal}).
After expressing the temperature as function of the scale 
factor by solving
\beq
\frac{d \rho_R}{da}+4\frac{\rho_R}{a}=\frac{\Gamma_\phi \rho_\phi}{Ha}\,,
\eeq
we find that 
the thermally-produced number density is given by
\begin{align}
    &n^T(T_{\rm RH})  = \frac{2 \beta_1}{\alpha^3} \frac{\sqrt{3} \rho_{\rm RH}^{5/2}}{m_{3/2}^4 M_P^3} \ln \left(\frac{\sqrt{\rho_{\rm end}}}{\sqrt{\alpha} T_{\rm RH}^2 } \right) ~\nonumber \\
    &+\frac{4 \beta_2}{\sqrt{3} \alpha^{5/2}} \frac{\rho_{\rm RH}^2}{m_{3/2}^2 M_P^3}+ \frac{2 \beta_3}{\sqrt{3}} \frac{\rho_{\rm RH}^{3/2}}{M_P^3}~\nonumber \\
    &+\frac{4 \beta_4}{3\sqrt{3} \alpha^{3/2}} \frac{\rho_{\rm RH} m_{3/2}^2}{M_P^3} + \frac{\beta_5 }{\sqrt{3} \alpha} \frac{\sqrt{\rho_{\rm RH}}m_{3/2}^4}{M_P^3} \,.
    \label{Eq:nthermal}
\end{align}    
We note that in our computation, we assumed that $4m_{3/2}^2 \ll s$, where $s = (p_1 + p_2)^2$, which would approximately correspond to $m_{3/2} \lesssim T_{\rm RH}$, and we integrated Eq.~(\ref{Eq:boltzmann4}) between $a_{\rm end}$ and $a_{\rm RH}$. \footnote{As discussed below, when $\trh < m_{3/2}$, the integration is limited between $\aend$ and $a_{3/2}$, where the latter corresponds to the scale factor when $T = m_{3/2}$.}
Moreover, since $\beta_1 \simeq 3.6$ is greater than $\beta_{i=2...5}$, the 
first term dominates the 
production process for $m_{\frac{3}{2}}< \trh$.

Using Eq.~(\ref{eq:relicabund1}) for the the relic abundance, we obtain
\bea
    \Omega^T h^2 & = & 5.9 \times 10^6 \, \frac{n^T(T_{\rm RH})}{T_{\rm RH}^3} \frac{m_{3/2}}{1~{\rm GeV}}
    \nonumber
    \\
   & \simeq & 3.6 \times 10^{-5} \left( \frac{T_{\rm RH}}{10^{10}~ {\rm GeV}} \right)^7 \left(12+ \ln \left(\frac{10^{10} {\rm GeV}}{\trh} \right) \right) \nonumber \\
   && \times \left( \frac{{\rm EeV}}{m_{3/2}} \right)^3 \, ,
    \label{Oh2therm}
\eea
where, in the approximation, we considered only the first term in Eq.~(\ref{Eq:nthermal}).
In Fig.~\ref{Fig:thermalcond}, we show in blue
the constraint on the relic abundance in the $(m_{3/2}, \trh)$ plane  from raritrons produced gravitationally from the thermal plasma.
As expected, the relic density generated by the thermal source is negligible compared with that generated by inflaton oscillations in all of the parameter space. This result is also valid for $k >2$.

\section{Gravitino dark matter}
\label{sec:gravdm}

The cosmological production of gravitinos has
been a constant source of potential cosmological problems due to its over-production. Standard thermal production (discussed briefly in Section \ref{sec:st-therm} below) sets upper limits to the reheating temperature after inflation 
\cite{nos,ehnos,kl,ekn,bbb,riotto,egnop,GMOP}.
There is also a non-thermal contribution to gravitino production when supersymmetry is broken during inflation \cite{riotto,Kallosh:1999jj,Giudice:1999yt,Kallosh:2000ve,Nilles:2001ry,Nilles:2001fg,Ema:2016oxl}. In these cases, as we will see as well below, it is the goldstino which is produced (i.e., the longitudinal component, rather than the transverse component), and this is typically the inflatino (the superpartner of the inflaton) during and immediately after inflation. Indeed it was argued that the longitudinal component of the gravitino at low energy may be unrelated to the inflatino produced after inflation \cite{Nilles:2001ry,Nilles:2001fg}. Whether or not the production of inflatinos is problematic is a model dependent question and inflatino production may even be kinematically suppressed \cite{Nilles:2001my}. Non-thermal production may also occur if the gravitino sound speed vanishes \cite{Benakli:2014bpa,hase,Kolb:2021xfn, Hashiba:2022bzi}.
However in this case, unless the models are constrained by eliminating the pseudoscalar, fermionic, and auxiliary components of the inflaton, no catastrophic production occurs \cite{Dudas:2021njv,Terada:2021rtp,Antoniadis:2021jtg}. 

In the remainder of this section, we consider first toy models involving two Majorana fermions coupled to the inflaton. 
This is a (highly) simplified example of  the inflaton coupling to the gravitino and inflatino. In this case, as in the non-thermal production of the raritron, the longitudinal component of the gravitino may be easily overproduced. However, as we just alluded, the produced state may not be the longitudinal component of the gravitino at low energy. Finally, we consider a specific model of inflation and supersymmetry breaking. If reheating is prolonged, the gravitino may be produced though with a suppressed abundance.

\subsection{Toy models}

In the previous section, we considered the gravitational production of raritrons from the inflaton condensate and thermal bath. As we have seen in Fig.~\ref{Fig:thermalcond}, the production from the condensate is dominant,
and from Fig.~\ref{Fig:rates},
we see that the production of the longitudinal (spin-$\frac12$)
component dominates over the transverse (spin-$\frac32$) component, 
particularly at low masses. In addition, 
as we will be interested in the production of gravitinos from the inflaton condensate as a concrete 
example in the next subsection, we would like to consider a toy model (not 
based on supergravity) which couples 
two spin-$\frac12$ Majorana fields, $\psi$ and $\chi$, to the inflaton. We 
will consider the production of $\psi$ through $\chi$--exchange having in mind 
the production of a goldstino through inflatino exchange when 
considering the supersymmetric analogue.

\begin{figure*}[t]
    \includegraphics[width=0.35\textwidth]{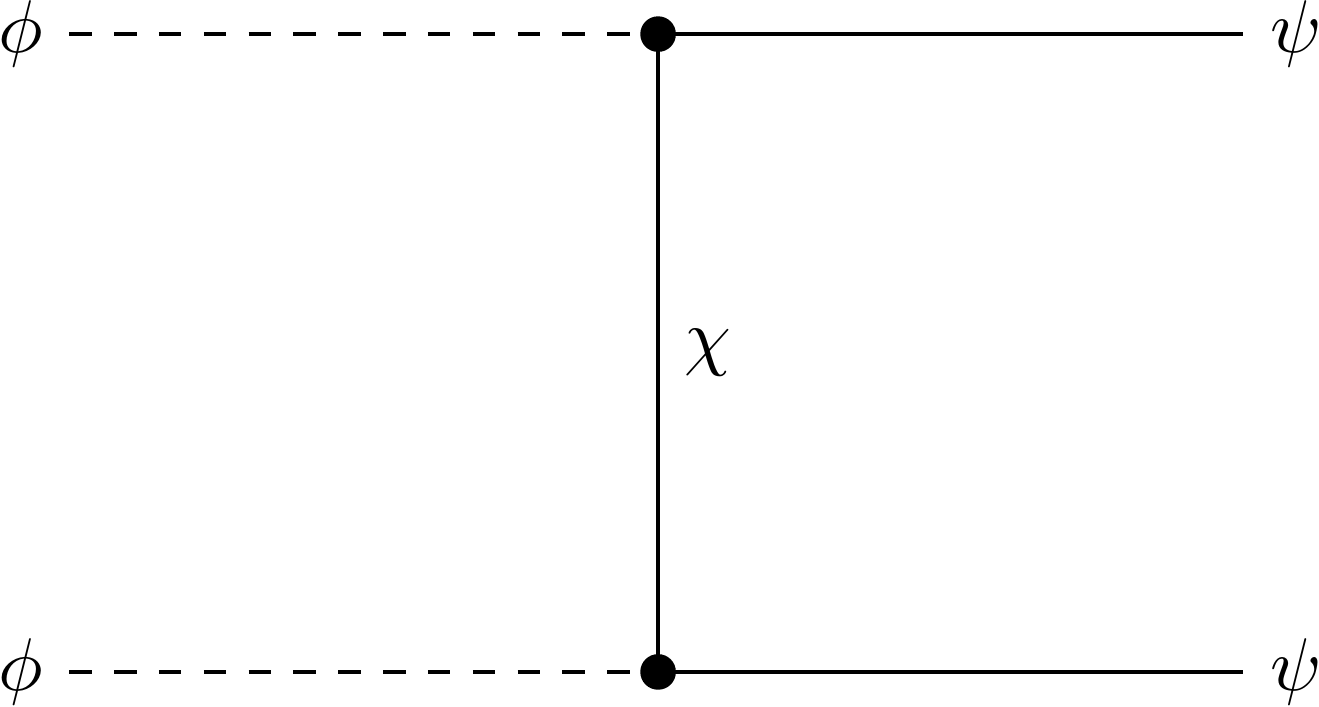}
    \hspace{2cm}
    \includegraphics[width=0.35\textwidth]{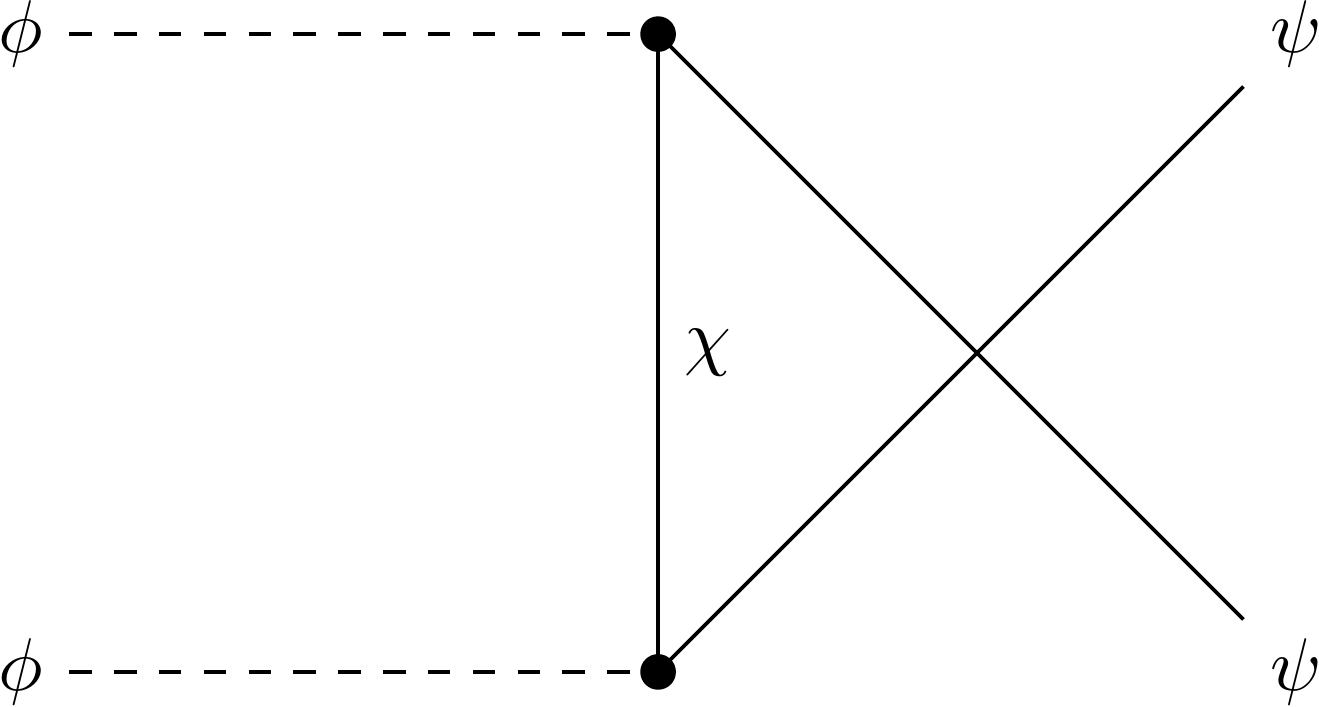}
    \caption{Feynman diagrams of the dark matter production processes.}
    \label{fig:feynman}
\end{figure*}

The toy model 
assumes a Yukawa coupling of the form 
\bea
{\cal L}_{\rm int} \; = \; -y\phi\bar\chi\psi + h.c. \, ,
\label{toy}
\eea
and a direct coupling of the inflaton to a pair of $\psi$'s is absent.
We further assume $m_\chi > m_\phi > m_\psi$ in our setup, so a direct decay of $\phi\to \chi\psi$ is not allowed kinematically.
The coherent oscillation of $\phi$ during reheating can however still produce $\psi$ through $\phi\phi \to \psi\psi$ by exchanging $\chi$, whose diagrams are shown in Fig.~\ref{fig:feynman}.

\begin{figure}[t]
    \centering
    \includegraphics[width=0.5\textwidth]{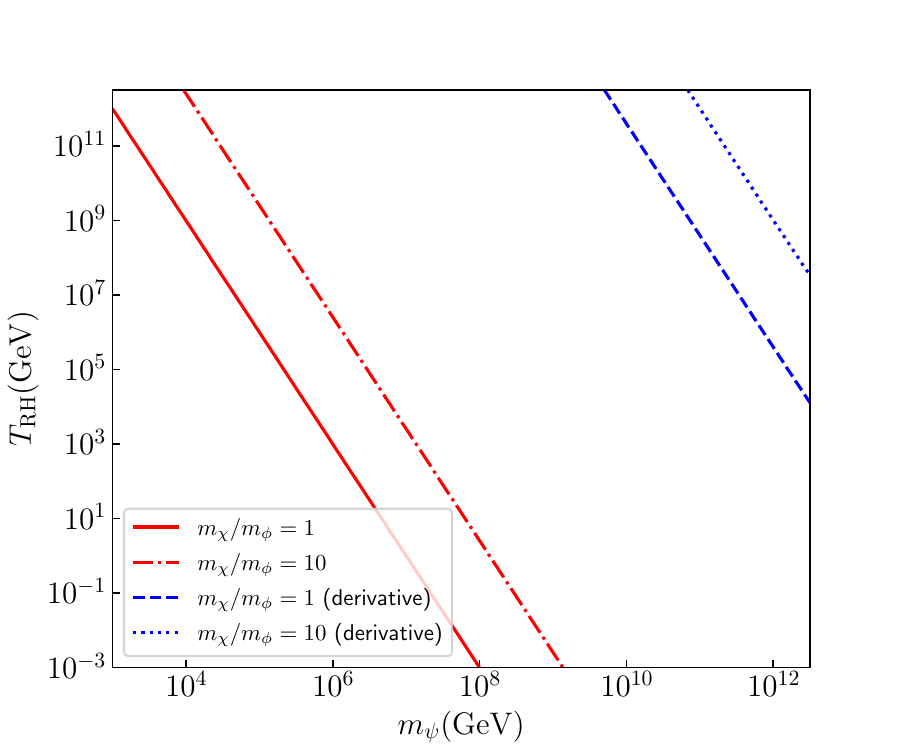}
    \caption{The $(m_{\psi},\trh)$ plane showing lines for $\Omega_\psi h^2=0.12$ with $m_\chi/m_\phi = 1$ and 10 as labeled as well as $y=1$, $m_\phi=1.7\times10^{13}$ GeV, and $\rho_{\rm end}=(5.2\times10^{15}~{\rm GeV})^4$. The red lines correspond to the relic density obtained from Eq.~(\ref{Opsi}) which was derived from the Lagrangian in Eq.~(\ref{toy}) (without the assumption $\tau_\chi \gg 1$). 
    The blue lines show the analogous result derived from the Lagrangian~(\ref{toy2}). 
    }
    \label{fig:Oh2_psi}
\end{figure}

As in the case of raritron production in the previous section, the abundance of $\psi$ can be obtained by integrating the Boltzmann equation given the production rate $\Gamma_{\phi\phi\to\psi\psi}\rho_\phi/m_\phi$
\bea
R(t) 
&=&
\frac{2 \times y^4}{\pi}\frac{\rho_\phi^2}{m_\phi^4}
\frac{\tau_\psi(1-\tau_\psi)^{3/2}}{(1+\tau_\chi-\tau_\psi)^2} \, ,
\label{nonsusyrate}
\eea
where $\tau_i\equiv m_i^2/m_\phi^2$.
As for previous rates, the factor of 2 in the numerator explicitly accounts for the fact that two $\psi$'s are produced per reaction. 
Using this rate, the Boltzmann equation (\ref{Eq:boltzmann4}) can be integrated to give 
\begin{eqnarray}
n(a_{\rm RH}) & = & \frac{4 y^4 M_P^3}{ \sqrt{3} \pi  m_\phi^4} \left(\frac{\rho_{\rm end}}{M_P^4} \right)^\frac12 \alpha T_{\rm RH}^4 \nonumber \\
&& \times \frac{\tau_\psi(1-\tau_\psi)^{3/2}}{(1+\tau_\chi-\tau_\psi)^2} \, ,
\label{npsi}
\end{eqnarray}
and 
\begin{widetext}
\bea
\Omega_{\psi}h^2 
&\simeq&
0.12 y^4\left(\frac{T_{\rm RH}}{10^{10}~{\rm GeV}}\right) \left(\frac{1.7\times10^{13}~{\rm GeV}} {m_\phi}\right)^2
\left( \frac{\rho_{\rm end}}{(5.2 \times 10^{15} {\rm GeV})^4} \right)^\frac12
\left(\frac{10^{14}~{\rm GeV}}{m_\chi}\right)^4\left(\frac{m_\psi}{33~{\rm TeV}}\right)^3 \, ,
\label{Opsi}
\eea
\end{widetext}
where we have taken the limit that $m_\chi \gg m_\phi \gg m_\psi$.
Fig.~\ref{fig:Oh2_psi} shows the parameter space satisfying $\Omega_\psi h^2=0.12$ with $y=1$, $m_\phi=1.7\times10^{13}$ GeV, and $\rho_{\rm end}=(5.2\times10^{15}~{\rm GeV})^4$. The results for this toy model is shown by the red lines with $m_\chi/m_\phi = 1$ (solid) and $m_\chi/m_\phi = 10$ (dot-dashed).
Indeed, because of the helicity suppression,
the relic density of $\psi$ scales as $m_\psi^3$ as opposed to $m_{3/2}^{-1}$ and hence we see very different behaviors when comparing the results in Fig.~\ref{Fig:thermalcond} and Fig.~\ref{fig:Oh2_psi}. 
For example, using the normalizations in Eq.~(\ref{Opsi}), $m_\chi/m_\phi \approx 5.9$ and $y=1$,  the spin-$\frac12$ fermion will provide the correct relic density when $m_\psi \simeq 33$ TeV. One further sees that rather than a divergence at small $m_\psi$, the relic density goes to 0, in this limit. This can be easily understood on the basis of helicity conservation.

We can also consider a similar toy model which matches more closely the supergravity couplings of the gravitino longitudinal mode. This simple Lagrangian  can be written as
\bea
{\cal L}_{\rm int} \; = \; -\frac{y}{M_P} \partial_\mu \phi\bar\chi \gamma^\mu \psi + h.c. \, .
\label{toy2}
\eea
Repeating the above exercise to calculate the production rate of $\psi$, we find
\bea
R(t) 
&=&
\frac{2 \times y^4}{\pi}\frac{\rho_\phi^2}{M_P^4}
\frac{\tau_\psi(1-\tau_\psi)^{3/2}}{(1+\tau_\chi-\tau_\psi)^2} \, ,
\eea
which is suppressed relative to the rate in Eq.~(\ref{nonsusyrate}) by a factor of $(m_\phi/M_P)^4$. The integration of the rate will be identical and the number density of $\psi$'s in Eq.~(\ref{npsi}) will be suppressed by the same factor. 
As a result, the mass needed to achieve $\Omega_\psi h^2 = 0.12$ is significantly larger, $m_\psi \simeq 2.4 \times 10^{11}$ GeV. The relation between $\trh$ and $m_\psi$ for the derivatively coupled toy model is also shown in Fig.~\ref{fig:Oh2_psi} (blue lines). As one can clearly see, the derivative coupling leading to the suppression requires a significantly 
larger mass, $m_\psi$ for a given reheating temperature in order to achieve the same relic density.

\subsection{Inflatino exchange}

We next turn to the example of the gravitino in supergravity models.
When supergravity models of inflation are considered, gravitino ($\psi_\mu$) generally couples to the inflaton ($\Phi$) and the inflatino ($\chi$) through the following terms:
\bea
{\cal L}_{\rm int} &=& -\frac{i}{\sqrt{2}M_P}\left[(\del_\mu\Phi)^*\bar\psi_\nu\gamma^\mu\gamma^\nu P_L\chi -(\del_\mu\Phi)\bar\chi P_R\gamma^\nu\gamma^\mu\psi_\nu\right].\nonumber\\
\label{eq:L_gravitino}
\eea
In the following argument, we assume that the imaginary part of $\Phi$ is strongly stabilized, and the canonically-normalized real part $\phi\equiv\sqrt{2}{\rm Re}\Phi$ is oscillating with an inflaton potential $V(\phi)$ after the end of inflation.

In addition to the gravitational production of gravitinos discussed previously, 
a pair of gravitinos can also be produced from  inflaton condensate via inflatino exchange.
There are both the t- and u-channels shown in Fig.~\ref{fig:feynman} with the replacement of $\psi \to \psi_\mu$.
Here, we are making the (naive) assumption, that the supersymmetry breaking sector is distinct from the inflationary sector and that the inflaton does not break supersymmetry. In this case, the goldstino (or spin-$\frac12$ component of the gravitino) is distinct from the inflatino. We return to a more realistic example in the next section. 
In the Boltzmann equation,
the production rate $R(t)$ is computed as
\bea
R(t) &=& \frac{2\rho_\phi^2}{9\pi M_P^4}\frac{(1-\tau_{3/2})^{7/2}}{\tau_{3/2}(1+\tau_\chi-\tau_{3/2})^2} \, ,
\label{susyrate}
\eea
where we have assumed $k=2$.
Note that as discussed in Appendix \ref{sec:ratecomp}, only the spin-$\frac12$ component of $\psi_\mu$ 
is produced. Details of the computation of the production rate are also given in Appendix \ref{sec:ratecomp}.
It is also interesting to see, by comparing 
equations (\ref{nonsusyrate}) and (\ref{susyrate}), that the 
production of the longitudinal component of $\psi_\mu$ (the spin-$\frac12$ part) is enhanced $\propto m_{3/2}^{-4}$
for light gravitino compared to 
the production of a spin-$\frac12$ 
fermion (\ref{nonsusyrate}) for reasons similar to those invoked 
when discussing the gravitational 
production of the 
raritrons.\footnote{Arguments based on 
the equivalence theorem can also be 
used to understand this.}

With this rate, the number density of gravitinos is 
obtained by solving Eq.~(\ref{Eq:boltzmann4}),
\begin{eqnarray}
n(a_{\rm RH}) & = & \frac{4 }{9 \sqrt{3} \pi  M_P} \left(\frac{\rho_{\rm end}}{M_P^4} \right)^\frac12 \alpha T_{\rm RH}^4 \nonumber \\
&& \times \frac{(1-\tau_{3/2})^{7/2}}{\tau_{3/2}(1+\tau_\chi-\tau_{3/2})^2} \, .
\label{Eq:ninflatino}
\end{eqnarray}
When $\tau_\chi \simeq 1$, and $\tau_{3/2} \ll 1$, this number density is roughly 8 times larger than that from graviton exchange given in Eq.~(\ref{n32tot}). 
It should not be surprising that the two results (\ref{n32tot}) 
and (\ref{Eq:ninflatino}) are so similar, since the exchange of a 
graviton involves couplings of the order $\partial_\mu/M_P$ generated by terms of the type $T_{\mu \nu}/M_P$, which have exactly the 
same form as the couplings between the inflaton, the 
inflatino and gravitino given in Eq. (\ref{eq:L_gravitino}). Only the graviton propagator differs from the inflatino propagator, but only in its structure, not in its order of magnitude.

The relic gravitino abundance can be then estimated by using Eq.~(\ref{eq:relicabund1})
\begin{eqnarray}
\Omega h^2 & \simeq & 2.4 \times 10^{10} \left( \frac{T_{\rm RH}}{10^{10} {\rm GeV}} \right) \left( \frac{\rho_{\rm end}}{(5.2 \times 10^{15} {\rm GeV})^4} \right)^\frac12 \nonumber \\ & & \left( \frac{m_\phi}{1.7 \times 10^{13} {\rm GeV}} \right)^2 \left( \frac{{\rm EeV}}{m_{3/2}} \right) \, ,
\label{oh2tot2}
\end{eqnarray}
which is, as expected, about 8 times larger than Eq.~(\ref{oh2tot}). 
Furthermore as we saw previously this abundance is highly dominated by the spin-$\frac12$ component. However, as we stressed earlier, this result ignores any contribution to supersymmetry breaking from the inflaton sector and any possible mixing between the spin-$\frac12$ partner of the  inflation, the inflatino, and the partner of the scalar associated with supersymmetry breaking in the vacuum.

\subsection{Specific Supergravity Model}
\label{sec:sugramodel}

Let us now consider a more realistic example, in which the identity of the goldstino evolves during the reheating process \cite{kmov,Dudas:2021njv}. 
To be more specific, we consider a model based on no-scale supergravity \cite{no-scale}. 
The K\"ahler potential can be written as
\bea
K = -3\ln\left[\Phi+\overline \Phi-\frac{1}{3}(|S|^2+|z|^2)+g(S,\overline S)+h(z,\overline z)\right] \, .\nonumber\\
\eea
The inflaton, $\phi$, is the real part of the canonically-normalized field, $\Phi \simeq \frac12 e^{\sqrt{2/3}\phi}$ (up to a small correction of order $\mu^2$ (defined below)).
The matter-like field $S$ 
and Polonyi field $z$ are stabilized by $g(S,\overline S)=|S|^4/\Lambda_S^2$ and $h(z,\overline z)=|z|^4/\Lambda_z^2$ \cite{ekn3,strongpol,dlmmo,ADinf,ego,dgmo,kmov,eno7,egno4,enov4,building}.
The inflaton can decay into gauge bosons and gauginos if the gauge kinetic function depends on the inflaton field value \cite{ekoty,klor,Terada:2014uia,dlmmo,egno4,building}. Barring a direct superpotential coupling of the inflaton to SM fields, this is the dominant decay mechanism in low-scale supersymmetric models. In models of high-scale supersymmetry, the inflaton  decays predominately into a pair of the SM Higgs bosons \cite{egno4,dgmo,dgkmo,KMO,kmov}.
The choice of the inflaton sector superpotential \cite{Cecotti}
\bea
W_{\rm inf} \; = \; \sqrt{3} m_\phi S(\Phi-1/2)
\eea
gives the Starobinsky-like inflaton potential \cite{eno7,FeKR,EGNO2}, and the inflaton energy density at the end of inflation becomes $\rho_{\rm end}=0.175m_\phi^2M_P^2$ with $m_\phi=3 \times10^{13}$ GeV  \cite{egno5,egnop}.
The inflatino is nearly degenerate in mass with inflaton. The scalar and fermionic components of $S$ are also nearly degenerate with the inflaton \cite{kmov}
(note that $\Lambda_S$ does not affect the spectrum at leading order).

One should, however, be careful about the time dependence of the mixing in the goldstino mode.
The goldstino in a non-static background is given by \cite{Nilles:2001fg} $\nu = G_I \chi^I + \slashed{\partial}\phi_I \chi^j G_J^I$, where $G\equiv K+\ln |W|^2$, $G_I\equiv \partial G/\partial \phi^I$, and  $G_J^I \equiv \partial G/\partial \phi^I \partial {\phi^*}^J$ with $\phi^I$ a superfield that participates in super-Higgs mechanism, and $\chi^I$ is the fermionic component in $\phi^I$.
We consider the Polonyi sector superpotential, given by \cite{pol}
\bea
W_P \; = \; \widetilde m(z+b) \, ,
\eea
with $b\simeq1/\sqrt3$.
When the reheating phase begins, 
the various contributions to the goldstino are given by
\bea
G_\Phi\simeq -\sqrt{\frac{3}{2}}\phi,~
G_S\simeq 2\mu +\sqrt{\frac{3}{2}}\frac{\phi}{\mu},\nonumber\\
G_z\simeq \sqrt{3}-\frac{3}{\sqrt2}\phi ,~
\slashed{\partial}\Phi  G_\Phi^\Phi \simeq m_\phi \phi \, ,
\eea
where $\mu\equiv \widetilde m/m_\phi\ll 1$.

Initially, when $\phi \gg \mu M_P$, supersymmetry is broken by the F-term of $S$. 
As discussed earlier, for the quadratic case considered here ($k=2$), the energy density of the inflaton scales as $\frac12 m_\phi^2 \phi^2 \propto a^{-3}$ and $\phi \propto a^{-\frac32}$ during reheating. 
But when $\phi/\mu \lesssim 1$, the primary component of the goldstino becomes the fermionic partner of the Polonyi field, $z$. 
We can estimate the corresponding scale factor $a_{\rm p}$, when $\phi/\mu = 1$ using
$\phi_{\rm p} = \phi_{\rm end} (\aend/a_{\rm p})^{3/2}$, and $\phi_{\rm end} = \sqrt{2 \rhoend}/m_\phi$, then 
\beq
\frac{a_{\rm p}}{a_e} = \left( \frac{\sqrt{2 \rho_e}}{M_P {\tilde m}} \right)^\frac23 \, ,
\eeq
and
\beq
\frac{a_{\rm p}}{a_{\rm RH}} = \left( \frac{2 \rho_{\rm RH}}{M_P^2 {\tilde m}^2} \right)^\frac13 \, .
\eeq

It would be tempting to deduce that for 
$a>a_{\rm p}$, the Polonyi field dominates in the goldstino and is produced by the inflaton condensate. However, this ignores the mixing
between the states. Furthermore, the degree of mixing \cite{Nilles:2001fg}, $\Delta$, gives rise to the gravitino sound speed,  $c_s^2 = 1- \Delta^2$, which can be expressed as \cite{Kolb:2021xfn}
\beq
\Delta^2 = \frac{4}{
\left(\vert\dot{\varphi}\vert^2 + \vert F \vert^2 \right)^2} \, 
\left\{ 
 |\dot{\varphi}|^2  |F|^2  - 
\left\vert \dot{\varphi}  \cdot F^*  \right\vert^2 \right\}  \;, 
\label{vs2}
\eeq
where $F^i \equiv {\rm e}^{K/2} K^{ij^*} (W_j + K_j W)$, and the $D$-term is absent in our analysis. The dot operator in Eq.~(\ref{vs2}) denotes a  scalar product with the K\"ahler metric $K_{ij}$, namely $\vert \dot{\varphi} \vert^2 = \dot{\varphi}^i \, K_{ij^*} \, \dot{\varphi}^{j*}$, and analogously for the other terms. 
As noted earlier if the gravitino sound speed vanishes \cite{Benakli:2014bpa,hase,Kolb:2021xfn}, catastrophic production of gravitinos ensues. Likewise, in the absence of mixing, divergent (as $m_{3/2} \to 0$) production
of gravitinos ensues as seen in Eqs.~(\ref{Eq:ninflatino}) and (\ref{oh2tot2}).
In the absence of constraints (for example the imposition of nilpotent fields), mixing is sufficiently large so as to suppress this non-thermal source of gravitino production \cite{Dudas:2021njv,Terada:2021rtp,Antoniadis:2021jtg}. Indeed the three-field model considered here, was also considered in \cite{Dudas:2021njv}. There, it was found that although the leading contribution to the sound speed may be small, the mixing parameter, $\Delta$, in this case is large. The detailed numerical analysis in a two-field model \cite{Nilles:2001fg} showed that the primary consequence of the mixing is that even though supersymmetry is initially broken ($a < a_{\rm p}$) by the inflationary sector and later ($a > a_{\rm p})$ through the Polonyi sector, the eigenstates rotate and the heavy mass eigenstate associated with the inflatino is always the field which is predominantly produced. Though a full numerical analysis of the three-field model was not performed, it was concluded that due to the large mixing, there is no catastrophic production of gravitinos in this model. 

It is interesting to note that the Lagrangian (\ref{eq:L_gravitino}), suitably extended to include the coupling of the Polonyi field to the gravitino,
can be separated into the parts providing the couplings of the transverse and longitudinal components. The latter will contain the mixing between the ``flavor'' eigenstates. This Lagrangian, written in \cite{Nilles:2001fg}, contains the basic elements found in our toy Lagrangian in Eq.~(\ref{toy2}). In agreement with the numerical results found in \cite{Nilles:2001fg}, our calculation of the production of $\psi$ is suppressed for low $m_\psi$.

In the remainder of this section, we briefly review the standard thermal production of gravitinos.

\subsection{Thermal production}
\label{sec:st-therm}

Before concluding this section, we can compare the production mechanisms above with the well known thermal production of gravitinos \cite{ekn,bbb,egnop,GMOP}.
So long as the scale of supersymmetry breaking is below the inflationary scale,
gravitinos can be singly produced, for example, by the scattering of two gluons producing a gluino and gravitino. Here, we use the parametrization in \cite{egnop,GMOP}
and consider only the gauge boson contribution to the production cross section, which we 
approximate as
\beq
\langle \sigma v \rangle \simeq \frac{26.24}{M_P^2}  \left(1+0.56\,\frac{m_{1/2}^2}{m_{3/2}^2}\right) \, .
\eeq
For $m_{3/2}$ significantly less than an assumed universal (at the GUT scale) gaugino mass, $m_{1/2}$, the 2nd term corresponding to the production of the longitudinal mode dominates. 
The production rate can then be written as
\beq
R_1 \simeq 0.4 \frac{T^6}{M_P^2}\left(1+0.56\,\frac{m_{1/2}^2}{m_{3/2}^2}\right) \, .
\eeq
Integrating this rate (for $k=2$) we arrive at
\beq
n(a_{\rm RH}) = \frac{4 \sqrt{3}}{9 \sqrt{\alpha}} \frac{0.4}{M_P} \trh^4 \left(1+0.56\,\frac{m_{1/2}^2}{m_{3/2}^2}\right) \,,
\eeq
and 
\beq
\Omega h^2 \simeq 0.04 \left(\frac{\trh}{10^{10}{\rm GeV}} \right) \left( \frac{m_{3/2}}{100~{\rm GeV}} \right)\left(1+0.56\,\frac{m_{1/2}^2}{m_{3/2}^2}\right)\,,
\label{oh2tot3}
\eeq
which gives the typical upper limit on the reheating temperature in supersymmetric models of $\trh \lesssim 2 \times 10^{10}$ GeV, for $m_{1/2} \simeq m_{3/2} = 100$ GeV, and the limit becomes stronger when $m_{1/2} > m_{3/2}$.
This limit results from the fact that the gravitino mass is related to the gaugino mass in a specific framework of SUSY breaking. 
We have neglected a kinematic factor which roughly requires $\trh \gtrsim m_{3/2}$.
Note that for these models we are using $g_{\rm RH} = 915/4$. 

In the case of high-scale supersymmetry breaking, gravitinos can only be pair produced if the masses of all other supersymmetric partners are greater than the inflationary scale. Nevertheless, gravitinos can be produced from SM particle annihilations with a rate
\beq
R_2 = \frac{T^{12}}{\Lambda^8} \, ,
\eeq
where $\Lambda^8\equiv(9/21.65)m_{3/2}^4M_P^4$ \cite{Benakli:2017whb,grav2,KMO}.
Integrating this rate gives
\beq
n(a_{\rm RH}) = \frac{21.65 }{9\sqrt{3} \sqrt{\alpha}} \frac{\trh^{10}}{m_{3/2}^4 M_P^3} \ln \frac{\rhoend}{\rhorh} \,,
\eeq
and
\bea
\Omega_{3/2}h^2&\simeq& 4 \times 10^{-6} \left(\frac{\rm EeV}{m_{3/2}}\right)^3\left(\frac{\trh}{10^{10}~{\rm GeV}}\right)^7
\nonumber \\ && \times \left(12+ \ln \left(\frac{10^{10} {\rm GeV}}{\trh} \right) \right) \,.
\label{oh2tot4}
\eea

\begin{figure}[!ht]
    \centering
    \includegraphics[width=0.52\textwidth]{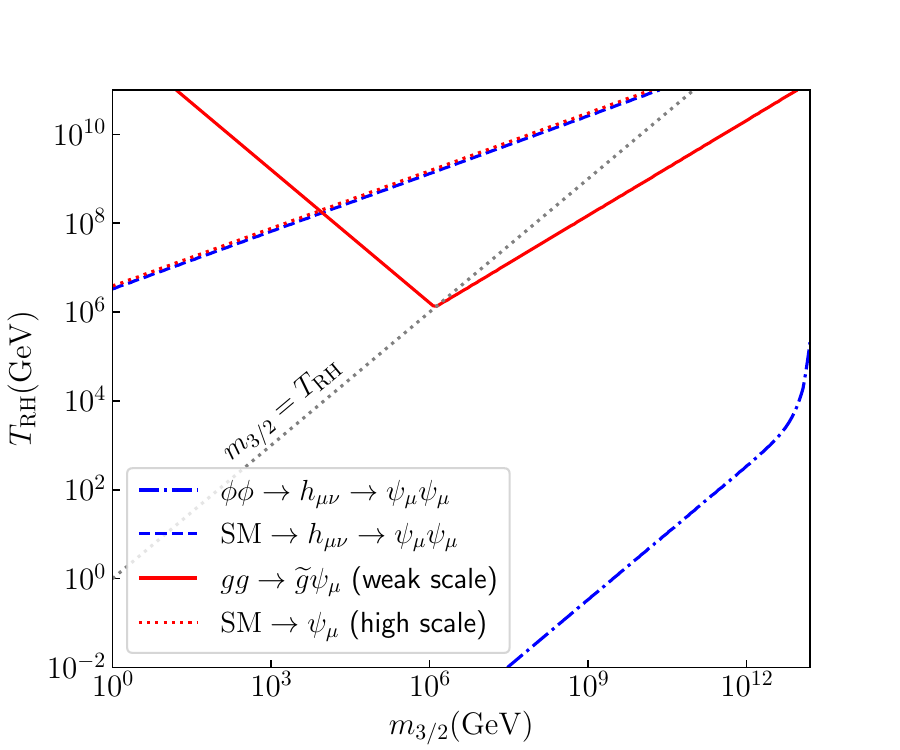}
    \caption{The $(m_{3/2},\trh)$ plane showing the contours of $\Omega_{3/2}h^2=0.12$.
    The blue dot-dashed line is derived from the inflaton condensate via single graviton exchange, Eq.~(\ref{oh2tot}).
    The blue dashed line is the thermal contribution from graviton exchange, given by Eq.~(\ref{Oh2therm}).
    The red solid line corresponds to single gravitino production when the scale of supersymmetry breaking is below the inflationary scale, Eq.~(\ref{oh2tot3}), where $m_{1/2}=m_{3/2}$ is assumed.
    The red dotted line corresponds to the case of high-scale supersymmetry, Eq.~(\ref{oh2tot4}), where gravitinos must be pair produced.
    }
    \label{fig:Oh2_32}
\end{figure}

In Fig.~\ref{fig:Oh2_32}, we show the regions of the parameter space allowed by the 
relic abundance constraint in the ($m_{3/2}, \trh$) plane for four of the 
processes we discussed in this paper. 
Specifically, we compare the thermal production of gravitinos in both low- and high-scale supersymmetric models from Eqs.~(\ref{oh2tot3}) and (\ref{oh2tot4}) with the non-thermal raritron production given in Eq.~(\ref{oh2tot}) and the thermal production from Eq.~(\ref{Oh2therm}).
The solid red line shows the ``classic" production of gravitinos in weak-scale supersymmetry, whose source is predominantly the gluons of the thermal 
bath. This is given by Eq.~(\ref{oh2tot3}), for which $\Omega h^2 \propto \trh m_{3/2}$. This provides the well known bound on the reheating temperature for stable gravitinos\footnote{A similar bound applies when gravitinos are unstable if R-parity is conserved as the produced gravitino abundance is transferred to the lightest supersymmetric particle.} and is applicable when $\trh > m_{3/2}$ \cite{GKMO2}.
For large gravitino masses, we must cut off the integration of the Boltzmann equation at $a_{3/2}$ corresponding to the scale factor when $T = m_{3/2}$, rather than integrating down to $\arh$.
For large masses, this leads to a suppression by a factor of $(a_{3/2}/\arh)^{9/4} = (\trh/m_{3/2})^6$. Thus for parameters with $\trh < m_{3/2}$, $\Omega h^2 \propto \trh^7 m_{3/2}^{-5}$.
This effect accounts for the change in the slope 
when $m_{3/2} \gtrsim \trh$ seen in the figure.  

This thermal bound on $\trh$ is greatly relaxed in models of high scale supersymmetry (shown here by the red dotted line)  as single production of gravitinos becomes kinematically forbidden \cite{Benakli:2017whb,grav2}. In this case, from Eq.~(\ref{oh2tot4}), we see that $\Omega h^2 \propto \trh^7 m_{3/2}^3$.
In contrast, the gravitational production of a stable spin-$\frac32$ raritron, whose 
source is the inflaton, provides a significantly stronger constraint, particularly at low masses. This constraint is shown by the blue dot-dashed line and given by Eq.~(\ref{oh2tot}) where $\Omega h^2 \propto \trh m_{3/2}^{-1}$. As discussed earlier the gravitational production of raritrons from the thermal bath is always sub-dominant.
It is shown by the blue dashed line from Eq.~(\ref{Oh2therm}) and is found extremely close to the line corresponding to the
thermal production in high-scale supersymmetry. 

This figure is one of the most 
important results of our studies, 
and admirably reflects the dominance 
of gravitational effects over 
classical thermal gravitino-raritron 
production, within the parameter 
space allowed by the unitarity 
limit, i.e., $m_{3/2}\gtrsim 40$ EeV (non-supersymmetric). However, we caution the reader that the thermal constraints shown here reflect the production of the gravitino in supersymmetric models. The gravitational production of the raritron is definitely not the gravitino in supersymmetry. As we have seen the gravitational production from the inflaton condensate produces primarily the longitudinal component and occurs just after inflation
when supersymmetry is broken by the inflationary sector. As a result the longitudinal component is the inflatino
and the resulting particles produced are not related to the gravitino at low energies.

\section{Summary}
\label{sec:sum}

Gravitational particle production after inflation is inevitable. All particles couple to gravity through their energy-momentum tensor and can be produced directly from the inflaton condensate during reheating. While it is difficult to create the thermal bath directly from minimal gravitational interactions \cite{cmov,cmo,Haque:2022kez,Barman:2022qgt}, the production of stable particles making up all or some of the dark matter is feasible \cite{MO}. 

While all particles couple to the inflaton through gravity, they do not couple equally. The production rate, $R$, for particles produced from the condensate are generally proportional to $\rho_\phi^2$. Since $\rho_\phi$ redshifts with the expansion of the Universe as in Eq.~(\ref{rhophia}), and the production rate redshifts faster than the Hubble rate, production occurs at the start of the reheating process. The production of scalars, $S$, is to a good approximation independent of the mass of the scalars when $m_S \ll m_\phi$ \cite{MO,cmov} and absent for massless vectors. However, the production rate for fermions is suppressed due to the necessity of a spin flip for the final state fermion \cite{cmov}. In this work, we considered the gravitational production of a massive spin-$\frac32$ particle dubbed the raritron. We do not however, necessarily associate this particle with the gravitino.
As in the case of scalars and spin-$\frac12$ fermions, the inflaton couples to raritrons through their respective energy-momentum tensors, $T^{\mu \nu}_{0}$ and $T_{3/2}^{\mu\nu}$. 

\begin{widetext}

\begin{figure}[!ht]
    \centering
    \includegraphics[width=0.42\textwidth]{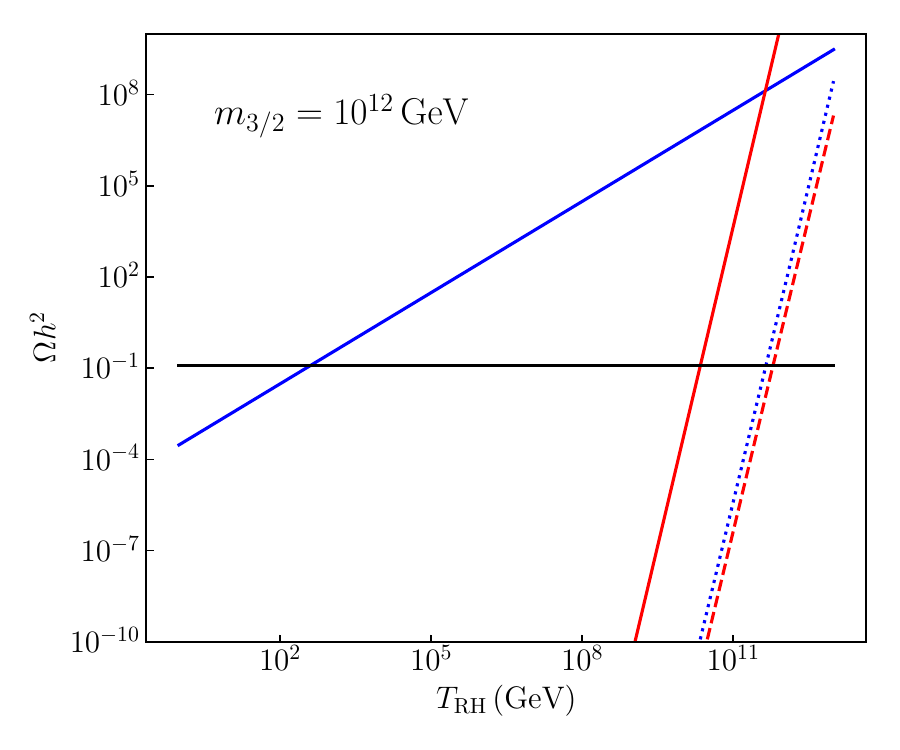}
    \includegraphics[width=0.42\textwidth]{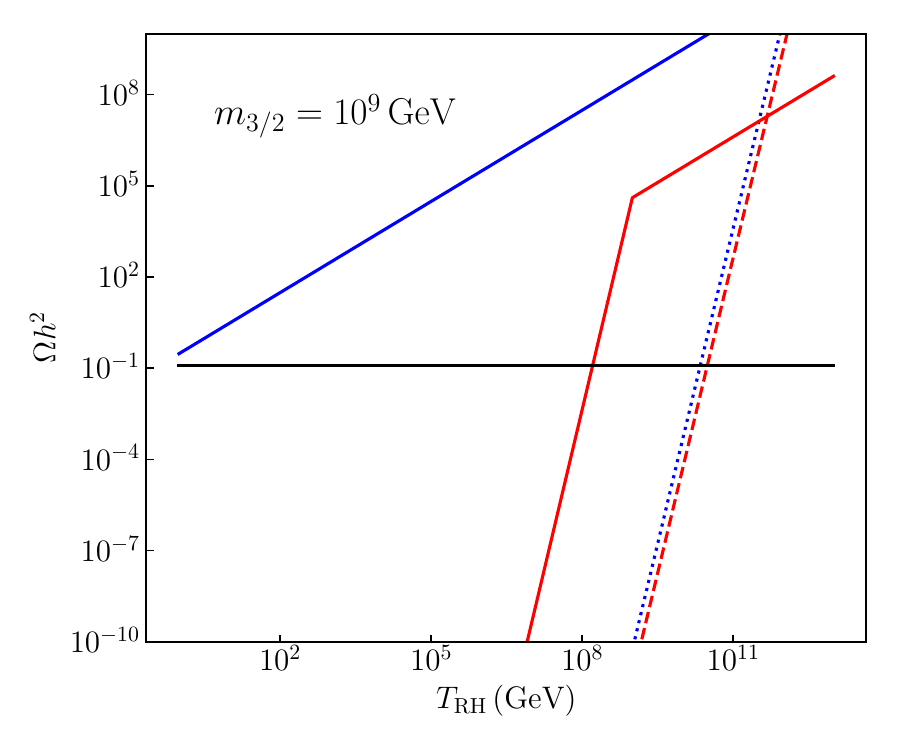}
    \includegraphics[width=0.42\textwidth]{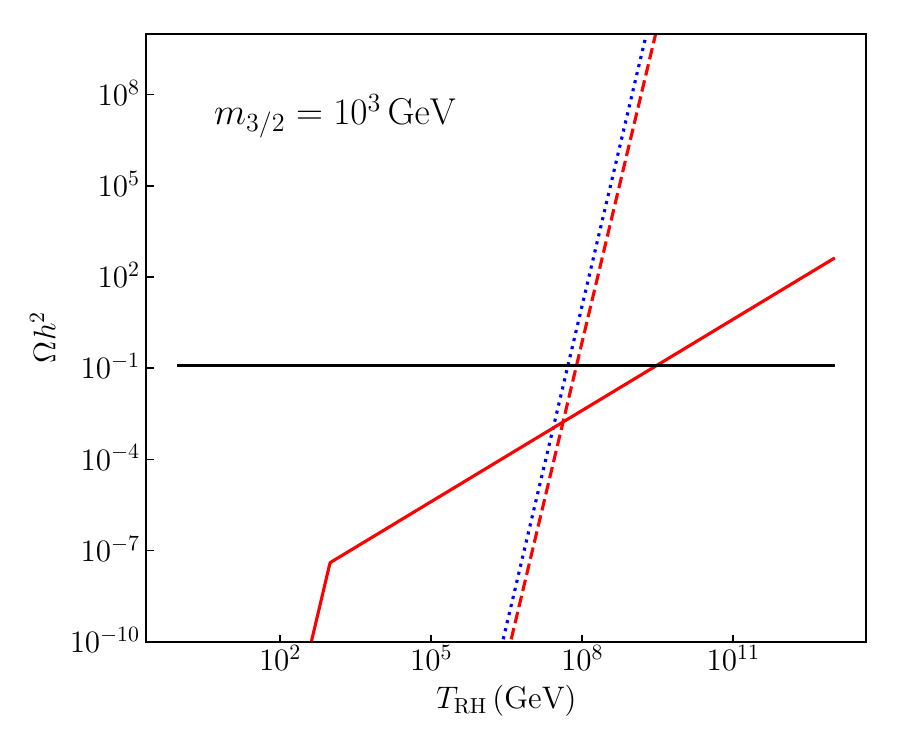}
    \caption{The relic abundance of raritrons/gravitinos, $\Omega h^2$ as a function of the reheating temperature for fixed $m_{3/2} = 1$ ZeV (upper left panel), $m_{3/2} = 1$ EeV (upper right panel),  and 1 TeV (lower panel). The blue solid line is derived from Eq.~(\ref{oh2tot}) and does not appear in the lower panel due to concerns over unitarity violations. The thermal production of raritrons mediated by gravity is shown as the blue dotted line from Eq.~(\ref{Oh2therm}).
   Also shown is the thermal production of gravitinos in both the case of weak-scale (solid red) and high-scale (dotted red) supersymmetry (with $m_{1/2}=m_{3/2}$) taken from Eqs.~(\ref{oh2tot3}) and (\ref{oh2tot4}) respectively. 
The horizontal black line at $\Omega h^2 = 0.12$ is shown for reference.
    }
\label{fig:Oh2_all_model}
\end{figure}

\end{widetext}

As we saw in Fig.~\ref{Fig:rates}, the production rate of raritrons is largely dominated by the production of the raritron longitudinal modes, particularly at low raritron masses, as the rate is proportional to $m_{3/2}^{-2}$. This yields a very large abundance of raritrons and unless $m_{3/2}$ is relatively large, the reheating temperature is strongly constrained as we showed in Fig.~\ref{Fig:thermalcond}. This  result is summarized in Fig.~\ref{fig:Oh2_all_model} where we
show the relic abundance $\Omega h^2$ as a function of the reheating temperature for three choices of $m_{3/2}$. The solid blue line is derived from Eq.~(\ref{oh2tot}). It is not shown in the lower panel with $m_{3/2} = 1$ TeV, as we expect unitarity violations at low masses. We also see in the upper right panel, that even $m_{3/2} = 1$ EeV ($10^9$ GeV), would require $\trh \lesssim 0.1$ GeV to avoid overproduction. 
When $m_{3/2} = 1$ ZeV ($10^{12}$ GeV), as in the upper left panel, the reheating temperature may be as large as $\sim$ 300 GeV. 
The horizontal black line is set at $\Omega h^2 = 0.12$ to guide the eye.

Other mechanisms for raritron/gravitino production are also shown in Fig.~\ref{fig:Oh2_all_model}. 
Note the huge variation in the relic abundance obtained from the different mechanisms. 
As discussed above, the thermal production of raritrons mediated by gravity is always sub-dominant when compared to the direct production from the inflaton condensate. This source of production is shown by the blue dotted curve in Fig.~\ref{fig:Oh2_all_model} taken from Eq.~(\ref{Oh2therm}). Extrapolating to larger masses and reheating temperatures we can see, however, because of the steep dependence ($\Omega h^2 \propto \trh^7)$, there are regions where thermal production dominates, but $\Omega h^2$ is orders of magnitude too large in this case.

We also show in Fig.~\ref{fig:Oh2_all_model}, the thermal production of gravitinos in both the case of weak-scale (solid red) and high-scale (dotted red) supersymmetry taken from Eqs.~(\ref{oh2tot3}) and (\ref{oh2tot4}) respectively. The abundance of thermally-produced gravitinos in weak--scale scale supersymmetry is proportional to $\trh$ so long as $\trh > m_{3/2}$. 
As we have already seen in Fig.~\ref{fig:Oh2_32}, for large gravitino masses, we must cut off the integration of the Boltzmann equation at $a_{3/2}$ corresponding to the scale factor when $T = m_{3/2}$, rather than integrating down to $\arh$.
This leads to a suppression by a factor of $(\trh/m_{3/2})^6$. For $m_{3/2} = 1$ ZeV, and the parameter range shown in Fig.~\ref{fig:Oh2_all_model},
$\trh < m_{3/2}$ and the we see only the steeper slope. For the other two values of $m_{3/2}$ shown, we see the change in slope when $\trh = m_{3/2}$. 
For $m_{3/2} = 1$ TeV, EeV, ZeV, we have limits of $\trh \lesssim$ $2 \times 10^9$ GeV, $4 \times 10^7$ GeV, and $5\times 10^9$ GeV respectively
to avoid overproduction of gravitinos in weak-scale supersymmetry. 

For the case of high-scale supersymmetry, cutting off the integration in the Boltzmann equation only results in a change in the log term in Eq.~(\ref{oh2tot4}) resulting in a replacement of $\trh$
with $m_{3/2}$ for $\trh<m_{3/2}$. This change is unobservable on the scale shown in the figure. 
The same is true for the thermal production via gravity in Eq.~(\ref{Oh2therm}). Since the relic abundance in both cases is $\propto 1/m_{3/2}^3$, the heavier the dark matter, the larger the permitted range of $\trh$.  The limits in the high-scale supersymmetry cases for  $m_{3/2} = 1$ TeV, Eev, ZeV, are  $\trh \lesssim$ $8\times 10^7$ GeV, $3.1\times10^{10}$ GeV, $6.3 \times 10^{11}$ GeV respectively. 

We have not included the production of gravitinos from inflatino exchange as that production is suppressed due to mixing with inflatinos.  A quantitative
measure of the abundance in that case would require an analysis similar to what is done in \cite{Nilles:2001fg}.

Of course we do not know how dark the dark sector is. At its darkest, gravitational interactions may play a leading role in the production of dark matter. A generic Rarita-Schwinger is easily overproduced in the early Universe through its (minimal) gravitational coupling to the inflaton. We have derived strong limits on the raritron mass in this case, though depending on the detailed model, unitarity limits may be even stronger.  The gravitino in models of broken supersymmetry can also be produced gravitationally, however only the transverse components are produced as the longitudinal states are primarily composed of the inflatino.  In this case the standard thermal production of gravitinos still provides limits on its mass and the inflationary reheating temperature.

\vspace{0.5cm} 
\noindent
\acknowledgements
The authors thank K. Benakli, G. Casagrande, E. Dudas, A. Guillen, E. Kolb, and A. Long for extremely valuable discussions during the completion of our work. This project has received support from the European Union's Horizon 2020 research and innovation programme under the Marie Sklodowska-Curie grant agreement No 860881-HIDDeN, and the IN2P3 Master Projet UCMN. The work of K.K. was supported in part by JSPS KAKENHI No. 20H00160. The work of K.A.O.~was supported in part by DOE grant DE-SC0011842 at the University of Minnesota. The work of S.V. was supported in part by DOE grant DE-SC0022148.

\appendix

\section{Energy-momentum tensor of spin-3/2 field}
\label{sec:tmn32}
In this appendix, we provide a brief review of the computation for the energy-momentum tensor of a spin-$\frac{3}{2}$ particle. We begin with a theory that closely resembles $\mathcal{N}=1$ pure supergravity, where the spin-$\frac{3}{2}$ particle is a Majorana fermion known as the gravitino.

We begin by introducing the full action, which is the sum of the Einstein-Hilbert action and the Rarita-Schwinger action for the massive gravitino,
\begin{align}
    S &=
    \int d^4x({\cal L}_2 + {\cal L}_{3/2}) \, ,
\end{align}
where
\begin{align}
    {\cal L}_2 &= -\frac{M_P^2}{2}eR \, ,\\
    {\cal L}_{3/2} &=
    -\frac{1}{4}\epsilon^{\mu\nu\rho\sigma}\overline\psi_\mu\gamma_5\gamma_\nu \lrnabla_\rho\psi_\sigma - \frac{1}{4}em_{3/2}\overline\psi_\mu[\gamma^\mu,\gamma^\nu]\psi_\nu~\nonumber \\
    &=
    \frac{i}{4}e\overline\psi_\mu\gamma^{\mu\nu\rho}\lrnabla_\rho\psi_\nu - \frac{1}{4}em_{3/2}\overline\psi_\mu[\gamma^\mu,\gamma^\nu]\psi_\nu \, ,
    \label{eq:Majorana spin-3/2}
\end{align}
with the determinant of the frame field given by $\text{det}~e_\mu^a \equiv e$ and $A\lrnabla_\mu B\equiv A\rnabla_\mu B-A\lnabla_\mu B$.
The covariant derivative acting on the spin-$\frac{3}{2}$ field is defined as
\begin{align}
    \nabla_\mu \psi_\nu &\equiv \left(\partial_\mu + \frac{1}{4}\omega_{\mu ab}\gamma^{ab}\right)\psi_\nu \, ,\\
    \overline\psi_\nu \lnabla_\mu &= \overline\psi_\nu \left(\ld_\mu - \frac{1}{4}\omega_{\mu ab}\gamma^{ab}\right)\, , \\
    \gamma^{ab} &= \gamma^{[a}\gamma^{b]} = \frac{1}{2}[\gamma^a,\gamma^b] \, ,
\end{align}
and $\overline\psi_\nu\ld_\mu\equiv \partial_\mu\overline\psi_\nu$.
The frame field $e_\mu^a$ is related to the flat Minkowski metric as
\begin{align}
    g_{\mu\nu} &= e^a_\mu e^b_\nu \eta_{ab} \, ,\\
    \eta_{ab} &= \textrm{diag}(+1,-1,-1,-1) \, .
\end{align}
The curvature tensor is given by 
\begin{align}
    R_{\mu\nu}{}^{ab} &= \partial_\mu\omega_\nu{}^{ab} - \partial_\nu\omega_\mu{}^{ab}+\omega_\mu{}^{ac}\omega_{\nu c}{}^b - \omega_\nu{}^{ac}\omega_{\mu c}{}^b,\\
    R_{\mu\nu\rho\sigma} &= e_{a\rho}e_{b\sigma}R_{\mu\nu}{}^{ab} \,,\\
    R_{\mu\nu} &= R^\rho{}_{\mu\rho\nu}\,,\\
    R &= e^\mu_a e^\nu_b R_{\mu\nu}{}^{ab} = g^{\mu\nu}R_{\mu\nu} \,,
\end{align}
where $\omega_{\mu a b}$ is the spin connection, given by
\begin{eqnarray}
    \omega_{\mu a b}=\omega_{\mu a b}(e)+K_{\mu a b } \, .
    \label{eq:omega(e,psi)}
\end{eqnarray}
Here $K_{\mu \nu \rho }$ is the contorsion tensor and 
\begin{equation}
\omega_\mu^{a b}(e)=2 e^{\nu[a} \partial_{[\mu} e_{\nu]}^{b]}-e^{\nu[a} e^{b] \sigma} e_{\mu c} \partial_\nu e_\sigma^c
\end{equation} 
is the torsionless contribution. In the non-supersymmetric case, the contorsion can be set to zero. However, in supergravity,  $K_{\mu \nu \rho }$ is expressed as a combination of terms bilinear in $\psi_\mu$. We derive the energy-momentum tensor $T_{3/2}^{\mu\nu}$ by varying the \textit{total} Lagrangian ${\cal L}={\cal L}_2+{\cal L}_{3/2}$ with respect to the frame field $e$, and then expressing the result in the Minkowski limit $g_{ab} \rightarrow \eta_{ab}$. 

The energy-momentum tensor can be derived from the Einstein equation, where the terms other than the pure 
spin-$2$ contribution are grouped to define $T_{3/2}^{\mu\nu}$. In this context, two main approaches exist: the Palatini and the metric formalism, or the first- and second-order formalism in the context of supergravity. Since it is a quite involved task to compute the energy-momentum tensor using either method, we briefly discuss the distinctions between the two. 

In the first-order formalism, the parameters $e$, $\omega$, and $\psi$ (with Lorentz indices suppressed) are treated as independent variables when varying the action. The spin connection $\omega$ is subsequently expressed as a function of $e$ and $\psi$ by requiring $\delta S/\delta \omega=0$~\cite{Deser:1976eh}. The second-order formalism treats only $e$ and $\psi$ as independent variables, with $\omega$ chosen to ensure that supersymmetry~\cite{Freedman:1976xh} is preserved. This approach assumes Eq.~(\ref{eq:omega(e,psi)}) at the starting point. For a more detailed discussion on the first- and second-order formalisms, see Ref.~\cite{Freedman:2012zz}.

In the first-order formalism, the total Lagrangian is treated as a function of $e$, $\omega$, and $\psi$.
The solution to $\delta S/\delta\omega=0$ is given by Eq.~(\ref{eq:omega(e,psi)}), with $\omega=\omega(e,\psi)$.
We then solve the condition $\delta S/\delta e|_{\omega=\omega(e,\psi)}=0$ and find that the Einstein equation is given by
\begin{align}
\left.G_{\mu\nu}(e,\omega)\right|_{\omega=\omega(e,\psi)} =&\displaystyle
\frac{e^{-1}e_{\mu a}}{M_P}\left.\frac{\delta {\cal L}_{3/2}}{\delta e^a_\nu}\right|_{\omega=\omega(e,\psi)} \, ,
\label{eq:Einstein eq 1}\\
G_{\mu\nu}=&R_{\mu\nu}-\frac{1}{2}g_{\mu\nu}R \, .
\end{align}
We note that the Einstein tensor $G_{\mu\nu}$ on the left-hand side of Eq.~(\ref{eq:Einstein eq 1}) includes both $e$ and $\psi$, the latter due to the contorsion term in the spin connection. Therefore, to derive the correct energy-momentum tensor, the $\psi$-dependent terms must be move to the right-hand side.

On the other hand, in the second-order formalism, deriving the energy-momentum tensor is more straightforward. As the Lagrangian is treated as a function of $e$ and $\psi$, with Eq.~(\ref{eq:omega(e,psi)}) already applied to eliminate the explicit $\omega$ dependence, the Einstein equation is simply derived from $\delta {\cal L}/\delta e =0$, and 
$G_{\mu\nu}$ does not depend on $\psi$. Consequently, 
the symmetrized energy-momentum tensor is defined as
\begin{align}
    T_{3/2,\mu\nu} &= e^{-1}e_{(\mu a}\frac{\delta {\cal L}_{3/2}}{\delta e^{\nu)}_a} \, ,
\end{align}
where as before ${\cal L}_{3/2}$ is given by eliminating the $\omega$ dependence in the second-order formalism.
This approach allows for a direct computation to compute the gravitino energy-momentum tensor, given by
\begin{align}
    T_{3/2,\mu\nu}&=
    -\frac{i}{4}\overline\psi_\rho\gamma_{(\mu}\lrnabla_{\nu)}\psi^\rho
    +\frac{i}{2}\overline\psi_{(\nu}\gamma_{\mu)}\lrnabla_\rho\psi^\rho
    +\frac{i}{2}\overline\psi^\rho\gamma_{(\mu}\lrnabla_\rho\psi_{\nu)} ,
    \label{eq:T_mn (Majorana spin-3/2)}
\end{align}
where we have used the equation of motion and the gravitino constraints. We note that the above result does not depend on the gravitino mass. In the flat Minkowski limit, we can replace $\lrnabla\to\lrd$ and neglect the four-Fermi terms originating from the torsion contribution as they are not relevant for our analysis, which leads to Eq.~\eqref{tmn32}. 

If we do not assume supersymmetry, the spin-$\frac{3}{2}$ particle is not necessarily a Majorana fermion.
For a Dirac spin-$\frac{3}{2}$ particle, the Lagrangian is given by
\begin{align}
    {\cal L}_{3/2} &=
    -\frac{1}{2}\epsilon^{\mu\nu\rho\sigma}\overline\psi_\mu\gamma_5\gamma_\nu \lrnabla_\rho\psi_\sigma - \frac{1}{2}em_{3/2}\overline\psi_\mu[\gamma^\mu,\gamma^\nu]\psi_\nu,
\end{align}
and the energy-momentum tensor is given by
\begin{align}
    T_{3/2,\mu\nu}^{\rm (Dirac)} &=
    2T_{3/2,\mu\nu}^{\rm (Majorana)} \, ,
\end{align}
where $T_{3/2,\mu\nu}^{\rm (Majorana)}$ is given by Eq.~(\ref{eq:T_mn (Majorana spin-3/2)}).

\section{Amplitudes and Thermal Rates}
\label{sec:thermal}
In this appendix, we compute the thermal production rate of raritrons, $R_{\frac{3}{2}}^T$. We consider only the massless Standard Model particles in the initial state, which include scalars, fermions, and gauge bosons. The dark matter production rate for the process $\mathrm{SM} + \mathrm{SM} \rightarrow \psi + \psi$ is given by the general expression Eq.~(\ref{thrate}), where we assumed that $4m_{3/2}^2 \ll s$ and included a factor of two in the numerator to account that two dark matter particles are produced per scattering event. 

We express the squared amplitudes in terms of the Mandelstam variables $s$ and $t$, which are given by
\begin{equation}
    t \; = \; \frac{s}{2} \left(\sqrt{1 - \frac{4m_{3/2}^2}{s}} \cos{\theta_{13}} - 1 \right) + m_{3/2}^2 \, ,
\end{equation}
\begin{equation}
    s \; = \; 2E_1 E_2(1- \cos{\theta_{12}}) \, .
\end{equation}

The general squared amplitude for the thermal processes involving SM initial states is given by Eq.~(\ref{Eq:ampscat}), where we include $4$ degrees for 1 complex Higgs doublet, $12$ degrees for $8$ gluons and 4 electroweak bosons, and $45$ degrees for 6 (anti)quarks with 3 colors, 3 (anti)charged leptons and 3 neutrinos. We note that the squared amplitudes include the symmetry factors of both the initial and final states, and this is indicated with an overbar.

When summing over all polarizations, the total squared amplitude of the gravity-mediated scalar production of raritrons is given by
\begin{equation}
\begin{aligned}
&|\mathcal{\overline{M}}^{0 \frac{3}{2}}|^2=
 \frac{1}{72 m_{3/2}^4 M_P^4 s^2}\left\{-s^2\left(s+2 t-2 m_\phi^2\right)^2\times\right.\\&\left.\left[m_\phi^4-2 m_\phi^2 t+t(s+t)\right]-72 m_{3/2}^{12}+24 m_{3/2}^{10}(7 s+12 t)\right.\\&\left. -2 m_{3/2}^8\left[47 s^2+264 s t+216 t^2 +72m_\phi^4-12m_\phi^2(s+12t)\right]
 \right.\\&\left.-2m_{3/2}^6\left[s^3-122 s^2t -288s t^2-144t^3+12 m_\phi^4(7s-12t)\right.\right.\\&\left.\left.+m_\phi^2\left(288t^2+216st -62s^2\right)\right]+m_{3/2}^4\left[s^4-34 s^3 t-210 s^2 t^2 
\right.\right.\\&\left.\left. -240 s t^3-72 t^4-72m_\phi^8+24m_\phi^6(s+12t)\right.\right.\\&\left.\left.-18 m_\phi^4\left(s^2+16 s t+24 t^2\right)-4 m_\phi^2\left(s^3-35 s^2 t 
-126 s t^2 \right.\right.\right.\\&\left.\left.\left.
-72 t^3\right)\right]+m_{3/2}^2s\left[24 m_\phi^8+s^4+6 s^3 t+44 s^2 t^2+64 s t^3 \right.\right.\\&\left.\left.
+24 t^4-32 m_\phi^6(s+3 t)+16 m_\phi^4\left(2 s^2+8 s t+9 t^2\right)\right.\right.\\&\left.\left.- 
 2 m_\phi^2\left(5s^3+24 s^2 t+80 s t^2+48 t^3\right)
\right]
\right. \Bigr \} \, ,
\label{fullM2}
\end{aligned}
\end{equation}
where $m_{\phi}$ is the scalar mass and $m_{3/2}$ is the raritron mass. For the incoming SM Higgs bosons, we set $m_\phi=0$, and this expression simplifies to
\begin{equation}
\begin{aligned}
&|\mathcal{\overline{M}}^{0 \frac{3}{2}}|^2 =
\frac{1}{72 m_{3/2}^4 M_P^4 s^2}\left[-s^2 t(s+t)(s+2 t)^2-72 m_{3/2}^{12}\right. \\
& +24 m_{3/2}^{10}(7 s+12 t)-2 m_{3/2}^8\left(47 s^2+264 s t+216 t^2\right)\\
& +m_{3/2}^6\left(-2 s^3+244 s^2 t+576 s t^2+288 t^3\right)\\
& +m_{3/2}^4\left(s^4-34 s^3 t-210 s^2 t^2-240 s t^3-72 t^4\right)\\&\left.+m_{3/2}^2 s\left(s^4+6 s^3 t+44 s^2 t^2+64 s t^3+24 t^4\right)\right] \, . 
\end{aligned}
\end{equation}

Similarly, the matrix element squared for the gravity-mediated raritron production from {\em massless} fermions is given by
\begin{equation}
\begin{aligned}
&|\mathcal{\overline{M}}^{\frac{1}{2} \frac{3}{2}}|^2 =
\frac{1}{144 m_{3/2}^4 M_P^4 s^2} \left\{
576 m_{3/2}^{12}-768 m_{3/2}^{10} (s+3 t) \right.\\&\left.
+4 m_{3/2}^8 \left(137 s^2+768 s t+864 t^2\right)-8 m_{3/2}^6 \left(58 s^3+265 s^2 t \right.\right.\\&\left.\left.
+504 s t^2+288 t^3\right)+4 m_{3/2}^4 \left(13 s^4+197 s^3 t+513 s^2 t^2\right.\right.\\&\left.\left.
+480 s t^3+144 t^4\right)+4 m_{3/2}^2 s \left(2 s^4-31 s^3 t-106 s^2 t^2\right.\right.\\&\left.\left.
-128 s t^3-48 t^4\right)+s^2 \left(s^4+10 s^3 t+42 s^2 t^2 \right.\right.\\&\left.\left. +64 s t^3+32 t^4\right) \right\} \, ,
\label{M1232}
 \end{aligned}    
\end{equation}
and the production from {\it massless} gauge bosons is given by
\begin{equation}
\begin{aligned}
&|\mathcal{\overline{M}}^{1 \frac{3}{2}}|^2 =
\frac{1}{18 m_{3/2}^4 M_P^4 s^2} \left\{
-36 m_{3/2}^{12}+12 m_{3/2}^{10} (s+12 t) \right.\\&\left.
-4 m_{3/2}^8 \left(23 s^2+30 s t+54 t^2\right)+4 m_{3/2}^6 \left(3 s^3+53 s^2 t \right.\right.\\&\left.\left.
+54 s t^2+36 t^3\right)-m_{3/2}^4 \left(25 s^4+118 s^3 t+186 s^2 t^2  \right.\right.\\&\left.\left.
+120 s t^3+36 t^4\right)+2 m_{3/2}^2 s \left(3 s^4+7 s^3 t+16 s^2 t^2 \right.\right.\\&\left.\left.+16 s t^3+6 t^4\right)-s^2 t (s+t) \left(s^2+2 s t+2 t^2\right) \right\} \, .
\label{M132}
 \end{aligned}    
\end{equation}
By evaluating the integral, we find that the thermal production rate of raritrons can be written as 
\bea
&&
R^T_{3/2} \; = \; \beta_1\frac{T^{12}}{m_{3/2}^4 M_P^4} + \beta_2\frac{T^{10}}{m_{3/2}^2 M_P^4}
+\beta_3 \frac{T^{8}}{M_P^4} 
\nonumber
\\
&&
+ \beta_4 \frac{m_{3/2}^2 T^6}{M_P^4} + \beta_5 \frac{m_{3/2}^4 T^4}{M_P^4}  \, ,
\eea
where
\begin{equation}
    \beta_1 \; = \; \frac{205511 \pi ^7}{85730400} \, ,
\end{equation}
\begin{equation}
    \beta_2 \; = \; \frac{16453 \zeta (5)^2}{15 \pi ^5} \, ,
\end{equation}
\begin{equation}
    \beta_3 \; = \; -\frac{369149 \pi ^3 }{93312000} \, ,
\end{equation}
\begin{equation}
    \beta_4 \; = \; -\frac{8759 \zeta (3)^2}{1152 \pi^5 } \, ,
\end{equation}
\begin{equation}
    \beta_5 \; = \; -\frac{49}{5760  \pi} \, .
\end{equation}

\section{Computation of the gravitino production rate}
\label{sec:ratecomp}
The production rate of the gravitino can be derived from the energy transfer rate from the inflaton energy density $\rho_\phi$ to the gravitino sector. Using the equation of state parameter $w_\phi = p_{\phi}/\rho_{\phi}$ for the inflaton, the evolution of $\rho_\phi$ follows
\beq
\frac{d\rho_{\phi}}{dt} + 3H(1+w_{\phi})\rho_{\phi} \;=\; -(1+w_{\phi})\Gamma_{\phi}\rho_{\phi}\,, 
\eeq
where the right-hand side is given by the energy transfer per space-time volume (Vol$_4$) due to the inflaton decay or scattering processes to particles $A$ and $B$, defined as
\bea
(1+w_\phi)\Gamma_\phi\rho_\phi \; \equiv \; \frac{\Delta E}{{\rm Vol}_4} \,,
\eea
where
\bea
\Delta E 
&\equiv& \int\frac{d^3p_A}{(2\pi)^32p_A^0} \frac{d^3p_B}{(2\pi)^32p_B^0}(p_A^0+p_B^0)\nonumber\\
&\times&
\left|\frac{1}{n!}\biggl< {\rm f}\left|\left(i\int d^4x_1{\cal L}_{\rm int}\right)\cdots \left(i\int d^4x_n{\cal L}_{\rm int}\right)\right|0\biggr>\right|^2\nonumber, \\
\label{eq:DeltaE}
\eea
and ${\cal L}_{\rm int}$ is the interaction Lagrangian ~(see \cite{GKMO2} for more details).

We decompose the oscillating inflaton as $\phi(t)\simeq \phi_0(t){\cal P}(t)$, where ${\cal P}$ represents the rapidly oscillating component and $\phi_0$ is its envelope that slowly evolves (redshifts) with time.
In practice, $\phi_0$ can be taken as a constant quantity when computing a reaction that occurs over time scales much shorter than the change in $\phi_0$. Therefore, the fast oscillating component can be decomposed as 
\bea
{\cal P}(t) \; = \; \sum_{n=-\infty}^{\infty}{\cal P}_n e^{-in\omega t} \, ,
\eea
where $\omega$ is the frequency of the inflaton oscillation.

Using the interaction Lagrangian ${\cal L}_{\rm int}$, given by Eq.~(\ref{eq:L_gravitino}), the amplitudes for the $t$- and $u$-channels are given by
\begin{widetext}
\bea
{\cal M}^{(n,m)}_t &=&
\frac{1}{4M_P^2}\frac{nm{\cal P}_n{\cal P}_m}{t_n-m_\chi^2}(\omega\phi_0)^2\bar u_\mu(p_A)\slashed{\delta}\gamma^\mu(\slashed{p}_{\phi,n}-\slashed{p}_A)\gamma^\nu\slashed{\delta}P_Ru_\nu^c(p_B)\,,\\
{\cal M}^{(n,m)}_u &=&
\frac{1}{4M_P^2}\frac{nm{\cal P}_n{\cal P}_m}{u_n-m_\chi^2}(\omega\phi_0)^2\bar u_\mu(p_A)\slashed{\delta}\gamma^\mu(\slashed{p}_B-\slashed{p}_{\phi,n})\gamma^\nu\slashed{\delta}P_Lu_\nu^c(p_B) \,,
\eea
\end{widetext}
where $p_{\phi,n}^\mu=(n\omega,\vec 0)^\mu$, $t_n\equiv(p_{\phi,n}-p_A)^2$, $u_n\equiv(p_B-p_{\phi,n})^2$, and $\slashed{\delta}\equiv \delta^0_\mu\gamma^\mu$ is introduced to account for $\del_\mu\phi=(\dot\phi,\vec 0)_\mu$.
Using these amplitudes, the energy transfer rate can be written as
\begin{widetext}
\bea
\frac{\Delta E}{{\rm Vol}_4} &=&
\int\frac{d^3p_A}{(2\pi)^32p_A^0} \frac{d^3p_B}{(2\pi)^32p_B^0}(p_A^0+p_B^0)\sum_{n+m>0}\sum_{\rm spin}|{\cal M}_t^{(n,m)}+{\cal M}_u^{(n,m)}|^2(2\pi)^4\delta^4(p_{\phi,n}+p_{\phi,m}-p_A-p_B) \, .
\eea
\end{widetext}
If we use the equation of motion for $\psi_\mu$, the sum of the amplitudes greatly simplifies and becomes
\begin{align}
{\cal M}_t^{(n,m)}+{\cal M}_u^{(n,m)}& =
\frac{m_{3/2}}{M_P^2}\frac{nm{\cal P}_n{\cal P}_m}{nm\omega^2+m_\chi^2-m_{3/2}^2}\nonumber\\
&
\times(\omega\phi_0)^2\delta^\mu_0\delta^\nu_0\bar u_\mu(p_A)u_\nu^c(p_B) \, ,
\label{eq:TotalAmplitude}
\end{align}
where we used $t_n=u_n=m_{3/2}^2-nm\omega^2$ and $p_A^0=p_B^0=(n+m)\omega/2$. We emphasize that only the $\mu=0$ contribution of $\psi_\mu$ is produced.
The gravitino wave function may be written as $\psi_\mu\sim \psi \epsilon_\mu$, where $\psi$ and $\epsilon_\mu$ denote the spin-$\frac12$ and 1 components, respectively. It is important to note that the spin $\pm$3/2 component is proportional to the transverse polarization of $\epsilon_\mu$, which does not have the $\mu=0$ component. As a result, only spin $\pm$1/2 mode of the gravitino may have a nonzero amplitude.

Thus, the amplitude given by Eq.~(\ref{eq:TotalAmplitude}) can be further simplified by substituting $u_0(p)=\sqrt{2/3}\epsilon_0(p)u(p)=\sqrt{2/3}(|\vec p|/m_{3/2})u(p)$, where $u(p)$ is the spin-$\frac12$ component that satisfies $(\slashed{p}-m_{3/2})u(p)=0$. 
Without specifying the oscillatory solution of the inflaton, we derive obtain the energy transfer rate:
\begin{widetext}
\bea
(1+w_\phi)\Gamma_{\phi\phi\to \psi_\mu\psi_\mu}\rho_\phi=\sum_{n+m\geq1}\frac{(nm)^2(n+m)^7({\cal P}_n{\cal P}_m)^2\omega^{11}\phi_0^4}{144\pi m_{3/2}^2M_P^4(nm\omega^2+m_\chi^2-m_{3/2}^2)^2}\left(1-\frac{4m_{3/2}^2}{(n+m)^2\omega^2}\right)^{7/2} \, .
\eea
\end{widetext}
For concreteness, we consider $V(\phi)=(m_\phi^2/2)\phi^2$, which implies that $w_\phi=0$.
The solution for $\phi$ can be expressed as 
$\phi(t)\simeq \phi_0(t)\cos(\omega t)$, where $\omega=m_\phi$ and $\rho_\phi\simeq (m_\phi^2/2)\phi_0^2$.
Consequently, ${\cal P}_{n=\pm1}=1/2$, and is zero otherwise. Therefore, only the $n=m=1$ modes contribute, and we obtain
\bea
\Gamma_{\phi\phi\to\psi_\mu\psi_\mu}&=&
\frac{2m_\phi\rho_\phi}{9\pi M_P^4}\frac{(1-\tau_{3/2})^{7/2}}{\tau_{3/2}(1+\tau_\chi-\tau_{3/2})^2} \, .
\eea
where $\tau_i\equiv m_i^2/m_\phi^2$.

\end{document}